\documentclass[superscriptaddress,aps,prb,noeprint,reprint,floatfix]{revtex4-2}
\usepackage{graphicx}
\usepackage{braket}
\usepackage{amssymb}
\usepackage{amsmath}
\usepackage{dcolumn}
\usepackage[utf8]{inputenc}
\usepackage{bm}
\usepackage{color}
\usepackage{xcolor}
\usepackage{float}

\begin{document}
\title{Quantum bicritical point and phase separation in a frustrated Heisenberg ladder}

\author{D . S. Almeida}
\author{R. R. Montenegro-Filho}
\affiliation{Laborat\'{o}rio de F\'{i}sica Te\'{o}rica e Computacional, Departamento de F\'{i}sica, Universidade Federal de Pernambuco, 50760-901 Recife-PE, Brasil}
\date{\today}

\begin{abstract}
We use the density matrix renormalization group (DMRG) and a hard-core boson map to investigate the quantum phase transitions present in the phase diagram of the frustrated Heisenberg ladder in a magnetic field. 
The quantum bicritical point is observed at the end of a first-order transition line, which is at the meeting of the two second-order transition lines that bound the fully polarized plateau. The characterization of the bicritical point was done by a hard-core boson mapping of the low-energy excitations from the fully polarized phase, and through DMRG by studying the probability density of finding the rung spins in a singlet or a triplet state with zero spin component along the magnetic field. In particular, we give conditions for the exchange couplings for the presence of the first-order transition line and the bicritical point in the phase diagram. Moreover, we unveil the phase-separated states for magnetization values inside the magnetization jump, and, in particular, the dependence on the system size of the energy curve as a function of magnetization. 
Finite-size scaling analysis of the transverse spin correlation functions has been used to estimate the critical points of the Kosterlitz-Thouless transitions from the fractional magnetization plateau $m=1/2$ to the respective gapless Luttinger liquid phases for some sets of parameters.
\end{abstract}

\maketitle

\section{Introduction}

The theory of phase transitions, in the classical \cite{chaikin2000principles} and quantum domains \cite{sachdev2011quantum,continentino_2017}, provides
a simple framework to understand the complex phenomenology
of systems with many interacting degrees of freedom. Paradigms of strongly correlated states of matter and
phase transitions are usually found in models and compounds
of magnetic systems. In particular,  
frustrated systems \cite{lacroix2011introduction,Vojta2018} are very fruitful to the physical investigation, presenting features such as
the quantum analog to the critical point of water \cite{Jimenez2021},
critical endpoints \cite{Stapmanns2018}; and
quantum bicritical points, 
as in the heavy-fermion metamagnet YbAgGe \cite{Tokiwa2013}. 
The bicritical point, specifically, is the end point of a first-order transition line, and, in its vicinity,  
the system is effectively described by two competing order
parameters \cite{Fisher1974,Kosterlitz1976,Morice2017,Lopes2020,Lopes2020a}.
In a first-order transition, the energy does not present a single global minimum and phase separation can occur. 
In a magnetic insulator, this
transition appears as a jump in the magnetization curve as a function of the magnetic field.

The spin-1/2 two-leg ladder \cite{Dagotto1999} is gapped, 
and the ground-state wavefunction is well described through short-range resonating-valence-bond states \cite{White1994}. In the presence
of a magnetic field \cite{giamarchi2003quantum}, the gap closes at a quantum critical point
due to the Zeeman effect. The gapless low-energy physics fits that of
the Luttinger liquid model \cite{PhysRevB.55.58,PhysRevB.59.11398} with a power-law decay of the correlation
functions \cite{Hikihara2001}.
In fact, the quasi-one-dimensional spin-1/2 ladder compound (C$_5$H$_{12}$N)$_2$CuBr$_4$ has been successfully
used to investigate quantum critical points and Luttinger
liquid physics \cite{Ruegg2008,Klanjsek2008,Thielemann2009}.
Furthermore, other interesting phenomena arise if frustration is added to the
model. Special features of frustrated two-leg ladders include their equivalence to spin-1 chains for some exchange patterns and magnetic field ranges \cite{Gelfand1991,White1996,Honecker_2000},
fractional magnetization plateaus \cite{Mila_1998,Tonegawa1998,Totsuka1998,Fouet2006,Penc2007,Michaud2010} and
first-order transitions \cite{Honecker_2000,Fouet2006,Michaud2010} (magnetization jumps) in their
magnetization curves, spinon and magnon condensation \cite{Fouet2006}, and Kosterlitz-Thouless  \cite{nobelkosterlitz,Kosterlitz1973} transition points \cite{Mila_1998,Totsuka1998,Tonegawa1998}.
Recently, the unfrustrated Ising ladder with four-spin interactions in a transverse magnetic field showed
phase coexistence \cite{Xavier2022}, and the frustrated model was used to understand mode splittings
in the ladder compound (C$_5$H$_{12}$N)$_2$CuBr$_4$ \cite{Nayak2020}. Furthermore, we mention that first-order \cite{PhysRevB.99.064404} and
Kosterlitz-Thouless transitions \cite{MontenegroFilho2020,Verissimo2019,Karlova2019,YamamotoPRB99} were identified
in one-dimensional ferrimagnetic models. 

Here, the density matrix renormalization group (DMRG) \cite{Schollw_ck_2005,PhysRevLett.69.2863,PhysRevB.48.10345} and a hard-core
boson mapping, taking as vacuum the fully polarized state, 
were used to investigate selected facets of the phase diagram of a two-leg ladder with
frustrated couplings along the two diagonals ($J_{\times}$) of the plaquettes in a magnetic field.
We characterize
the quantum bicritical point and associated first-order transition line, particularly
the phase-separated states inside the magnetization jump. Moreover, we present a careful
numerical estimate of the Kosterlitz-Thouless critical points at which the $m=1/2$ plateau
closes. We mention that the localization of Kosterlitz-Thouless transition
points using numerical methods is challenging, since the gap is exponentially
small near the critical point \cite{Gebhard2022}.  

In Sec. \ref{sec:model}, we show the Hamiltonian of the frustrated ladder, briefly discuss the relevant features of this model, and
present the numerical methods used in our calculations. The magnetization curves and a general description of the phase diagram
are provided in Sec. \ref{sec:phasediagram}. In Sec. \ref{sec:hcmapping}, we obtain the single-particle excitations from the fully
polarized state through a hard-core boson mapping. The bicritical point is identified in Sec. \ref{sec:hfphases} by analyzing the
transverse spin correlation functions and the density of dimer excitations in the competing gapless phases. The Kosterlitz-Thouless
transition points are estimated through the transverse spin correlation functions in Sec. \ref{sec:kttransitions}.
While we have considered a single set of model parameters in the above sections, in Sec. \ref{sec:othervalues} 
we exhibit the phase diagrams for two other sets to investigate the stability of the observed phases. A summary of the paper is shown in Sec. \ref{sec:sum}.

\section{Model and methods}
\label{sec:model}

\begin{figure}
    \centering
    \includegraphics[width=0.4\textwidth]{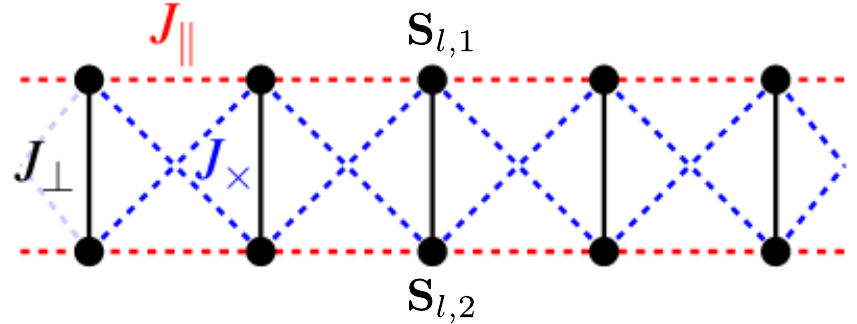}
    \caption{The frustrated ladder. We consider the phase diagram of the chain as a function of the magnetic field $h$ and 
    $J_{\times}$, fixing $J_{\perp}\equiv1$ and $J_{\parallel}$.}
    \label{fig:escada}
\end{figure}

The Hamiltonian of the frustrated two-leg spin-1/2 ladder with $L$ dimers and open boundary conditions is given by 
\begin{align}
     \mathcal{H}&=\sum_{l=1}^{L}\mathbf{S}_{l,1}\cdot \mathbf{S}_{l,2}+J_{\parallel}\sum_{l=1}^{L-1}\left(\mathbf{S}_{l,1}\cdot \mathbf{S}_{l+1,1}+\mathbf{S}_{l,2}\cdot \mathbf{S}_{l+1,2}\right) \notag \\
     &+J_{\times}\sum_{l=1}^{L-1}\left(\mathbf{S}_{l,1}\cdot \mathbf{S}_{l+1,2}+\mathbf{S}_{l+1,1}\cdot \mathbf{S}_{l,2}\right)\nonumber\\
     &-hS^z,
\label{eq:ham}
\end{align}
and is schematically illustrated in Fig.\ref{fig:escada}.
We consider the phase diagram of the chain as a function of $J_{\times}$ and the magnetic field $h$ 
for fixed values of the exchange couplings on the rungs, $J_{\perp}\equiv1$, and
along the legs, $J_{\parallel}$. The direction of the external magnetic field $h$ is
defined as the $z$-direction, while $S^z=\sum_l S^z_l=\sum_l(S^z_{l,1}+S^z_{l,2})$ is the $z$-component of the total spin,  and $g\mu_{B}\equiv1$. We notice, also, that the ladder is symmetric under the exchange of $J_\times$ and $J_\parallel$, and of the label of the spins in odd (or even) dimers. 

In the regime $J_{\perp}\gg (J_\parallel,J_\times)$ and $h>0$, the ladder has each dimer in a 
$S^z_l=1$ triplet or in a singlet \cite{Mila_1998,Fouet2006,Penc2007}. In this case, the Hamiltonian can be mapped onto \cite{Mila_1998} the XXZ Heisenberg model or the model of interacting spinless fermions in a linear chain. The magnetization per dimer $m$ as a function of $h$ exhibits plateaus
at $m=0$ and $m=1$, and a fractional plateau at $m=1/2$ for some combinations of the exchange couplings.
In particular, KT transitions are predicted \cite{Mila_1998} within this
approximation. Furthermore, the presence of magnetization jumps with DMRG was observed \cite{Fouet2006,Michaud2010}, and
for a range of values of $J_{\times}$ for $J_{\times}=J_{\parallel}$ using DMRG and Maxwell construction \cite{Honecker_2000}. 
Moreover, the phase diagram for $h=0$ shows a rung-singlet phase for $J_{\perp}$ dominant and 
a rung-triplet phase (Haldane phase) in the other regimes. A detailed discussion of
the $h=0$ case can be found in Ref. [\onlinecite{Wessel2017}]. 

We used the DMRG and exact diagonalization numerical methods to calculate the magnetization curves and spin correlations of finite-size systems.
In the case of DMRG, we used the codes provided by the Algorithms and Libraries for Physics Simulations (ALPS) project \cite{Bauer2011} and the ITensor library \cite{SciPost}, beyond our code. In the DMRG simulations, 
the maximum number of states kept per block, or bond dimension in the case of the tensor code,
was 1355, and we have considered chains with open boundaries. The typical discarded weight was less than $1\times 10^{-10}$, while the maximum discarded weight was $\sim 1\times 10^{-8}$. We also applied a hard-core boson mapping to examine the single-particle excitations from the fully polarized state.

\section{Magnetization curves and phase diagram}
\label{sec:phasediagram}
\begin{figure}
    \centering
    \includegraphics[width=0.48\textwidth]{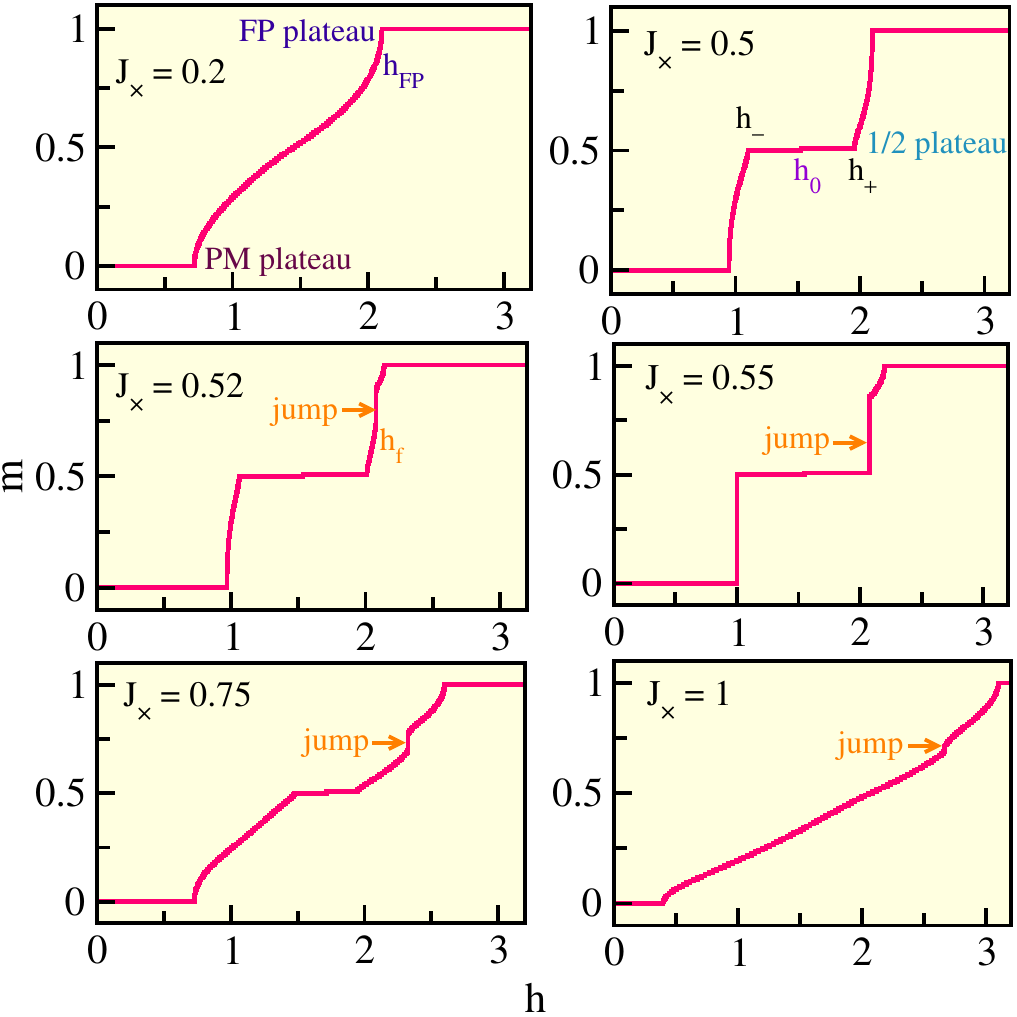}
    \caption{Magnetization per dimer $m=\langle S^z \rangle/L$ vs magnetic field $h$ for a ladder with $L=128$ dimers and $J_{\parallel} = 0.55$. The quantum paramagnetic (PM), 1/2, and fully polarized (FP) 
    plateaus are indicated, as well as the critical magnetic fields $h_{PM}$, $h_-$, $h_+$, and $h_{FP}$.
    We denote by $h_f$ the magnetic field at which the magnetization jump occurs.
    At the field $h_0$, the 1/2 plateau splits into two steps in the finite-size system with open boundary conditions.}
    \label{fig:tudo}
\end{figure}

We present in Fig. \ref{fig:tudo} the typical features of the magnetization per dimer $m$ as a function of $h$ as we change $J_{\times}$ 
keeping $J_{\parallel}(=0.55)$ fixed. 
For a finite-size system, $m(h)=\langle S^z \rangle/L$ have finite-size
steps of width $\Delta h(S^z)=h_{S^z+}-h_{S^z-}$ at a given value of $S^z$, with $h_{S^z\pm}=\pm \left[E(S^z\pm1)-E(S^z)\right]$ being the extreme points of these steps, where $E(S^z)$ is the total energy at a given value of $S^z$. Magnetization plateaus in the thermodynamic limit are characterized by
$\Delta h(S^z)\neq 0$ as $L \to \infty$.
\begin{figure}
    \centering
    \includegraphics[width=0.48\textwidth]{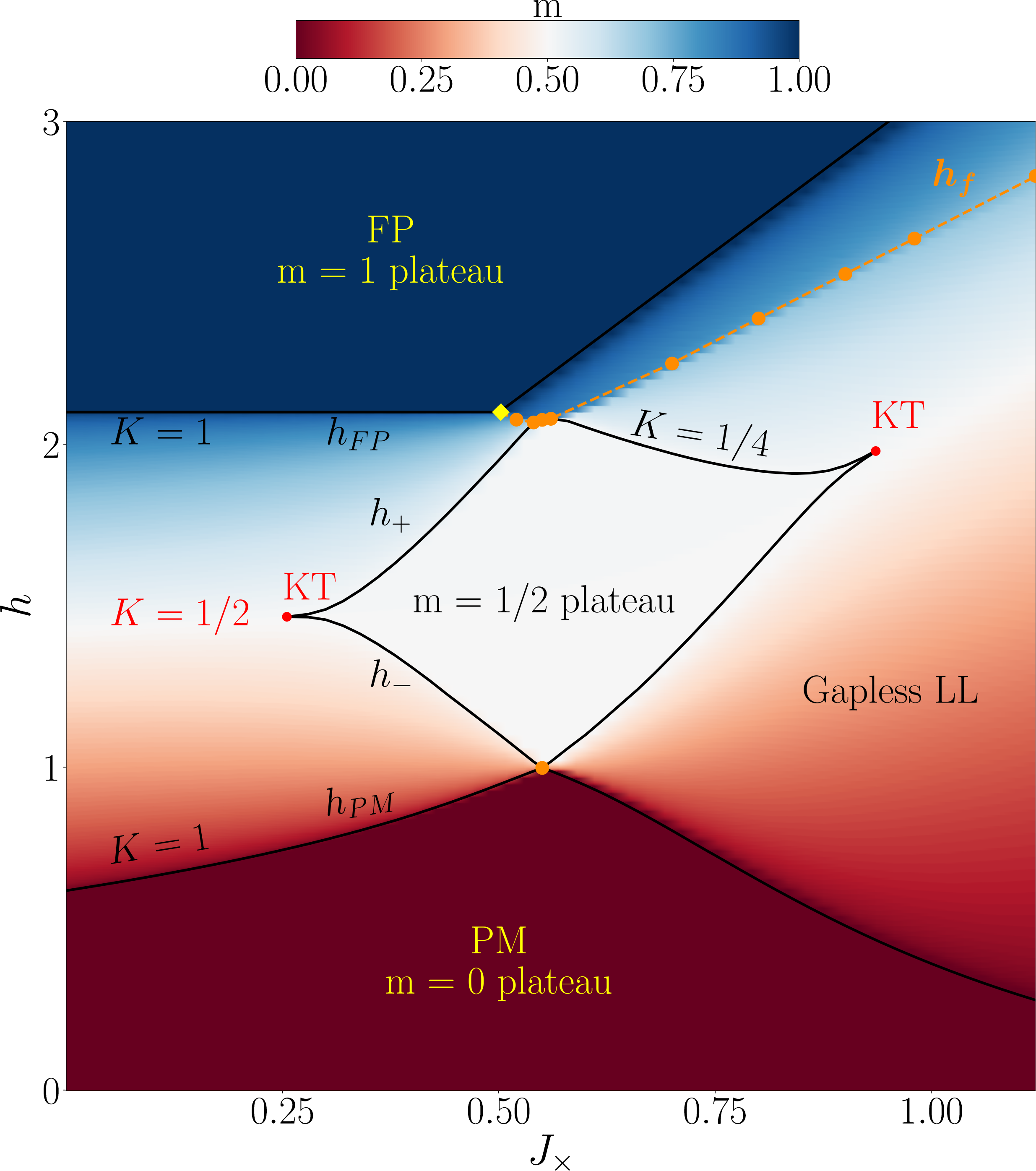}
    \caption{
    DMRG results for the thermodynamic-limit magnetic field $h$ versus frustration $J_{\times}$ phase diagram for $J_{\parallel}=0.55$. The color code gives the magnetization for a finite-size chain with $L=128$.
    The gapped fully polarized (FP), 1/2, and paramagnetic (PM) plateaus are bounded by the lines $h_{FP}$, $h_+$, $h_-$, and $h_{PM}$, respectively. 
    The intermediate regions between plateaus are gapless Luttinger liquid (LL)
    phases. We indicate the limiting value of the Luttinger liquid parameter $K$ as the second-order transition points are approached from the gapless phases. Along the first-order transition line $h_f$, the magnetization curve $m$ versus $h$ exhibits a jump. The line $h_f$ starts at a ($\color{yellow}\blacklozenge$) bicritical point in the $h_{FP}$ line. We mark with a ($\color{orange}\bullet$) the two first-order transition points on the line $J_{\times}=J_{\parallel}$. 
    The $m=1/2$ plateau closes at the ($\color{red}\bullet$) Kosterlitz-Thouless (KT) transition points, for which $K\rightarrow1/2$ on the gapless side of the transition.}
    \label{fig:fase}
\end{figure}

The Oshikawa-Yamanaka-Affleck argument \cite{oshikawa} states that a magnetization plateau can appear at the magnetization $m$ if 
\begin{equation}
(S_u-m_u)=\text{integer},
\end{equation}
where $S_u$ is the maximum total spin, and $m_u$ is the magnetization in the unit cell of the ground-state wavefunction. In our case,
the unit period of the Hamiltonian has two
spin-1/2 sites. Thus, magnetization plateaus that do not spontaneously break translation symmetry can occur at $m=0$, the paramagnetic (PM) plateau, and at $m=m_s=1$, the fully polarized (FP) plateau.
For any value of $J_{\times}$, these two magnetization plateaus are observed in Fig. \ref{fig:tudo}. Their critical fields are
defined as $h_{FP}$ for the FP plateau and $h_{PM}$ for the PM plateau. 
For moderate values of $J_{\times}$, a magnetization plateau appears at $1/2$ of the saturation magnetization.
In this case, the ground state has a doubled unit cell with four spin-1/2 sites.
The critical fields at the extreme of the 1/2 plateau are indicated by $h_-$ and $h_+$.
In a finite-size chain with open boundary conditions, this plateau is made of two steps joined at $h=h_0$. These two steps  consist in domain walls \cite{Fouet2006}, and we briefly discuss them in Sec. \ref{sec:kttransitions}.  
Between the thermodynamic-limit magnetization plateaus, there are gapless Luttinger liquid (LL) phases \cite{giamarchi2003quantum} with critical power-law transverse spin correlation functions with the asymptotic form 
\begin{equation}
 \Gamma(r)\sim \frac{1}{r^{1/2K}},
 \label{eq:ll-corr}
\end{equation}
where $K$ is the Luttinger liquid exponent and $r$ is the distance between spins along the chain.
Furthermore, a magnetization \textit{jump} at $h=h_f$ for $J_{\times}>0.5$.

In Fig. \ref{fig:fase}, we show our estimated phase diagram $h$ vs. $J_{\times}$ for the frustrated ladder with $J_{\parallel}=0.55$.
There, we highlight the magnetization $m$, the color code, for a system with $L=128$, and the thermodynamic-limit boundary lines of the magnetization plateaus. For fixed values of $J_{\times}$, there are typically second-order transitions at the extremes of the plateau: $h_{FP}$, $h_{-}$, $h_{+}$, and $h_{PM}$. 
As the second-order transition point is approached from the gapless side of the transition \cite{Cazalilla_2011}, 
$K\rightarrow 1$ at the boundaries of the plateaus with the same periodicity of the Hamiltonian (PM-plateau and FP-plateau), 
while $K\rightarrow 1/4$ at the boundaries of the 1/2-plateau.
However, at some points, the transitions in the plateau extremes are of the first-order kind: for $J_{\times}=0.55=J_{\parallel}$, at the PM-plateau and 1/2-plateau transition lines. 
Magnetization curves exhibit magnetization jumps in the first-order transition line $h_f$. 
In particular, this line ends at a bicritical point at $J_{\times}=0.5$, which is the 
endpoint of the two second-order transition lines that bound the FP plateau.
For fixed $m$, the 1/2 plateau closes at two points through transitions of Kosterlitz-Thouless (KT) type: $J_{\times, \text{KT}_1}=0.255\pm0.005$ and $h_{\text{KT}_1}=1.467\pm0.002$; $J_{\times, \text{KT}_2}=0.935\pm0.005$ and $h_{\text{KT}_2}=1.98\pm0.01$. In these cases, $K\rightarrow 1/2$ as the transition points are
approached from its gapless side.

In the following, we discuss some features of the phase diagram shown in Fig. \ref{fig:fase}. In particular, we estimate the first-order transition line for $J_{\parallel}=0.55$ and characterize the bicritical transition point at the end of this line. Moreover, we use a confident numerical methodology to estimate the exact critical values of $h$ and $J_{\times}$ for the KT transitions. Furthermore, we exhibit phase diagrams for $J_\parallel=0.2\text{ and }0.8$ to explore the stability of the observed phases and transition lines under changes in $J_\parallel$.

\section{Hard-core boson mapping}
\label{sec:hcmapping}

The critical transition line from the fully polarized state, $h_{\text{FP}}$, can be precisely determined
by mapping the spin variables to hard-core boson operators \cite{Zapf2014}: 
 \begin{align}
  S^z_{l,i}&=\frac{1}{2}-a_{l,i}^\dagger a_{l,i}=\frac{1}{2}-n_{l,i},\nonumber\\
  S^+_{l,i}&=a_{l,i},\nonumber\\
  S^-_{l,i}&=a^\dagger_{l,i}.\nonumber
 \end{align}
To preserve the momentum angular commutation relations, $[S^{+}_{l,i},S^{-}_{m,j}]=2\delta_{lm}\delta_{ij}S^z_{l,i}$ and $[S^z_{m,j},S^{\pm}_{l,i}]=\pm \delta_{lm}\delta_{ij}S^{\pm}_{l,i},$, with $S^{\pm}=S^x\pm iS^y$; besides the constraints $S^+_{l,i}S^-_{l,i}+S^-_{l,i}S^+_{l,i}=1$ and $(S^+_{l,i})^2=(S^-_{l,i})^2=0$ specific to spin-1/2 operators, the commutation relation  
\begin{equation}
[a_{l,i},a^\dagger_{m,j}]=\delta_{lm}\delta_{ij}(1-2n_{l,i})
\end{equation}
must be satisfied by the bosonic operators, in addition to $(a_{l,i})^2=(a_{l,i}^\dagger)^2=0$.

\begin{figure}
    \centering
    \includegraphics[width=0.48\textwidth]{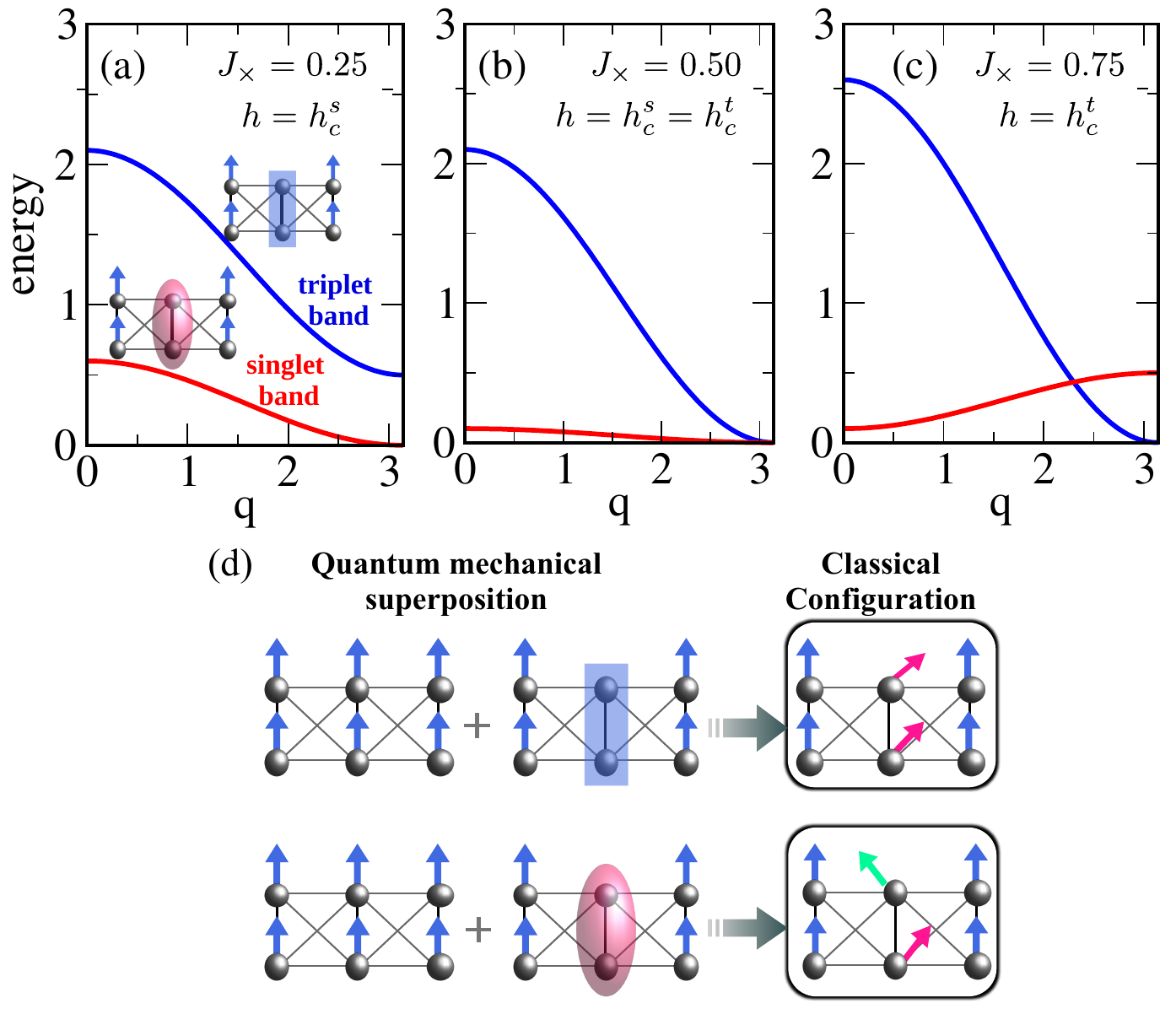}
    \caption{Free singlet and triplet hard-core bosons bands for $J_{\parallel}=0.55$ and the indicated 
    values of $J_{\times}$ and magnetic field $h$. The critical fields $h_c^s=1+2J_{\parallel}$ and 
    $h_c^t=1+2J_{\times}$. In (a), we illustrate with a blue box a triplet state 
    $\ket{t_{0,l}}=\frac{1}{\sqrt{2}}(\ket{\downarrow\uparrow}+\ket{\uparrow\downarrow})$ and with a red ellipse a 
    singlet state $\ket{s_l}=\frac{1}{\sqrt{2}}(\ket{\downarrow\uparrow}-\ket{\uparrow\downarrow})$ between the 
    spins of a dimer.} 
    \label{fig:freebands}
\end{figure}

In the bosonic variables, the Hamiltonian (\ref{eq:ham}) is written as
\begin{align}
 H^{\text{(free hc)}}&=(-\frac{1}{2}+J_{\parallel}+J_{\times})N+hN\nonumber\\
  &+\frac{1}{2}\sum_l(a^\dagger_{l,1} a_{l,2}+a^\dagger_{l,2} a_{l,1})\nonumber\\
  &+\frac{J_{\parallel}}{2}\sum_l\sum_{i=1}^2 (a^\dagger_{l,i} a_{l+1,i}+a^\dagger_{l+1,i} a_{l,i})\nonumber\\
  &+\frac{J_{\times}}{2}\sum_l(a^\dagger_{l,1} a_{l+1,2}+a^\dagger_{l,2} a_{l+1,1}+\text{H.c.}),
\label{eq:hamboson}
\end{align}
where $N=\sum_l(n_{l,1}+n_{l,2})$ is the total number of bosons, and we have discarded interaction and constant 
terms. The Hamiltonian (\ref{eq:hamboson}) can be straightforwardly diagonalized if we use odd and even combinations 
of $a_{l,1}$ and $a_{l,2}$:
\begin{align}
 s^\dagger_{l}&=\frac{1}{\sqrt{2}}(a_{l,1}^\dagger-a_{l,2}^\dagger)\nonumber\\
 t^\dagger_{0,l}&=\frac{1}{\sqrt{2}}(a_{l,1}^\dagger+a_{l,2}^\dagger).
 \label{eq:transf}
\end{align}
When applied to the fully polarized state $\ket{\text{FP}}$, $s^\dagger_{l}$ creates a singlet state between the two sites at the 
dimer $l$:
\begin{equation}
 s^\dagger_l\ket{\text{FP}}=\frac{1}{\sqrt{2}}(\ket{\downarrow\uparrow}_l-\ket{\uparrow\downarrow}_l)=\ket{s_l},
\end{equation}
while $t^\dagger_{0,l}$ creates a triplet state with total spin component $S^z_l=0$ at the dimer $l$:
\begin{equation}
 t^\dagger_{0,l}\ket{\text{FP}}=\frac{1}{\sqrt{2}}(\ket{\downarrow\uparrow}_l+\ket{\uparrow\downarrow}_l)=\ket{t_{0,l}}.
\end{equation}
Writing the
Hamiltonian (\ref{eq:hamboson}) in the variables (\ref{eq:transf}) and Fourier transforming, we arrive in the diagonal 
Hamiltonian: 
\begin{equation}
 H^{\text{(free hc)}}=\sum_q \varepsilon^{s}_q s^\dagger_q s_q+\sum_q \varepsilon^{t}_q t^\dagger_{0,q} t_{0,q},
\end{equation}
where the dispersion relations are given by
\begin{align}
 \varepsilon^{t}_q&=-(J_{\parallel}+J_{\times})+(J_{\parallel}+J_{\times})\cos(q)+h, \label{eq:dispersion1}\\
 \varepsilon^{s}_q&=-1-(J_{\parallel}+J_{\times})+(J_{\parallel}-J_{\times})\cos(q)+h, \label{eq:dispersion2}
\end{align}
and are shown in Fig. \ref{fig:freebands} for specific values of $J_{\times}$ and $h$, 
with $J_{\parallel}=0.55$.

The magnetic fields used in Fig. \ref{fig:freebands} are the critical fields $h_{\text{FP}}$ for
the respective values of $J_{\times}$ and are determined as follows. Since the lowest energy of the 
triplet band is 
\begin{equation}
\varepsilon^{t}_{\scriptsize{min}}=\varepsilon^{t}_{q=\pi}=-2(J_{\parallel}+J_{\times})+h, 
\label{eq:tripletmin}
\end{equation}
the triplet component condenses at a field 
\begin{equation}
h_c^t=2J_{\parallel}+2J_{\times}. 
\label{eq:hc}
\end{equation}
On the other hand, the  
lowest energy of the singlet band is
\begin{align}
\varepsilon^{s}_{\scriptsize{min}}&=\varepsilon^{s}_{q=\pi}=-1-2J_{\parallel}+h~\text{ for $J_{\times}<J_{\parallel}$; or}\label{eq:singletmin1}\\
\varepsilon^{s}_{\scriptsize{min}}&=\varepsilon^{s}_{q=0}=-1-2J_{\times}+h~\text{ for $J_{\times}>J_{\parallel}$.}\label{eq:singletmin2}
\end{align}
Thus, the singlet component condenses at 
\begin{align}
h_c^s&=1+2J_{\parallel}\text{ for $J_{\times}<J_{\parallel}$; or}\label{eq:hs1}\\ 
h_c^s&=1+2J_{\times}\text{ for $J_{\times}>J_{\parallel}$.}\label{eq:hs2}
\end{align}

\begin{figure}
    \centering
    \includegraphics[width=0.35\textwidth]{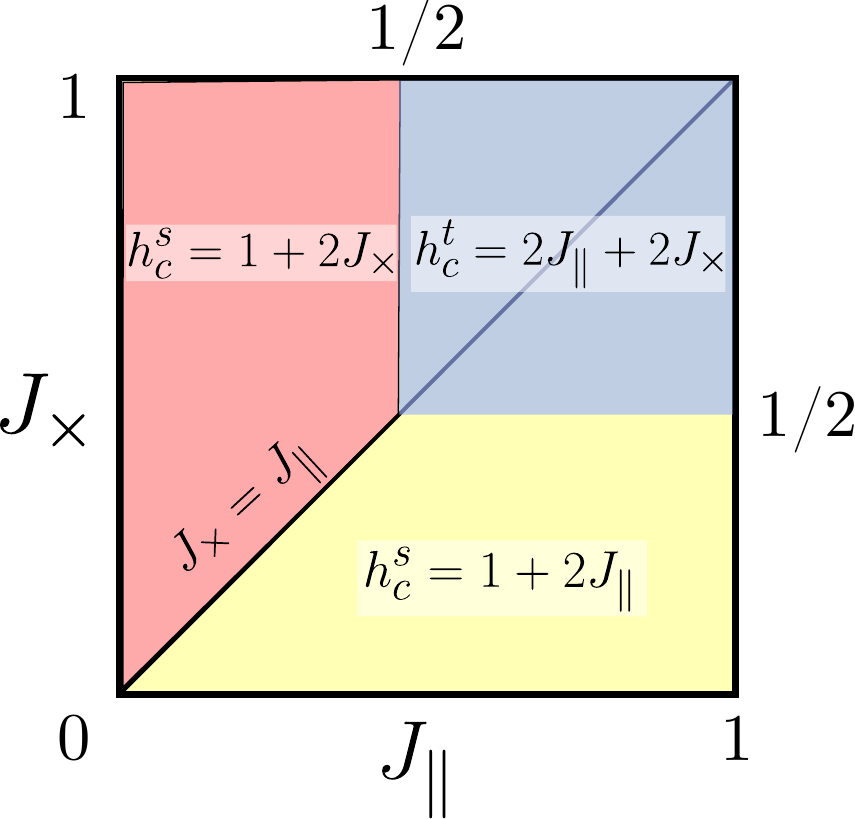}
    \caption{Expressions for the critical fully polarized field $h_{FP}$ as 
    a function of $J_\times$ and $J_\parallel$, as obtained by comparing  
    the minimum energy of the single-particle triplet and singlet bands: 
    Eqs. (\ref{eq:tripletmin}), (\ref{eq:singletmin1}), and (\ref{eq:singletmin2}).} 
    \label{fig:freehfp}
\end{figure}

We determine the critical field $h_{FP}$ for given values of $J_\times$ and $J_\parallel$ 
by finding the minimum energy between those in (\ref{eq:tripletmin}), 
(\ref{eq:singletmin1}), and (\ref{eq:singletmin2}). 
First, we draw the line $J_\times=J_\parallel$ in Fig. \ref{fig:freehfp}, since
the minimum of the singlet band is given by Eq. (\ref{eq:hs1}) for $J_\times<J_\parallel$, 
and according to Eq. (\ref{eq:hs2}) for $J_\times>J_\parallel$. 
In sector $J_\times<J_\parallel$, the minimum energy of the singlet
band crosses the minimum of the triplet one at $J_\times=1/2$, as indicated by a comparison
of Eqs. (\ref{eq:tripletmin}) and (\ref{eq:singletmin1}). In the sector $J_\times>J_\parallel$, 
Eqs. (\ref{eq:tripletmin}) and (\ref{eq:singletmin2}) show that
the crossing occurs at $J_\parallel=1/2$. For example,
consider the phase diagram of Fig. \ref{fig:fase}. Fixing
$J_\parallel$ at 0.55 and changing $J_\times$, we have
$h_{FP}=h^s_c=2.1$ for $J_\times<0.5$, and $h_{FP}=h^t_c=2.1+2J_\times$
for $J_\times>0.5$. From Fig. \ref{fig:freehfp}, we also note that the crossing between the singlet and triplet bands is absent for $J_\parallel<1/2$.

\section{Magnetic order in the high-field regime: bicritical point, first-order transition line, and phase separation}
\label{sec:hfphases}

Here, we show the average triplet and singlet components, as well as the correlation functions
calculated with DMRG for $J_\perp=0.55$. 
This section mainly aims at the region $m>1/2$.

The average probability density of singlets at the dimer $l$ can be written as 
\begin{align}
\braket{n^s_l}=\braket{s^\dagger_l s_l}&=\frac{1}{4}-\braket{\mathbf{S}_{l,1}\cdot\mathbf{S}_{l,2}}+\braket{n_{l,1}n_{l,2}}\nonumber\\
                      &\approx \frac{1}{4}-\braket{\mathbf{S}_{l,1}\cdot\mathbf{S}_{l,2}},
\label{eq:orderns}
\end{align}
while the average probability density of triplets $\ket{t_0}$ at the dimer $l$ as
\begin{align}
 \braket{n^{t_0}_l}=\braket{t^\dagger_{0,l} t_{0,l}}&=\frac{3}{4}-\braket{S^z_l}+\braket{\mathbf{S}_{l,1}\cdot\mathbf{S}_{l,2}}-\braket{n_{l,1}n_{l,2}}\nonumber\\
                                 &\approx \frac{3}{4}-\braket{S^z_l}+\braket{\mathbf{S}_{l,1}\cdot\mathbf{S}_{l,2}},
\label{eq:ordert0}
\end{align}
where $S^z_l=S^z_{l,1}+S^z_{l,2}$. In both expressions, we have discarded
the term $\braket{n_{l,1}n_{l,2}}$, which is nonzero if the dimer is in the triplet state
$\ket{\downarrow\downarrow}$, which has a low probability of occurrence. 
\begin{figure}
    \centering
    \includegraphics[width=0.43\textwidth]{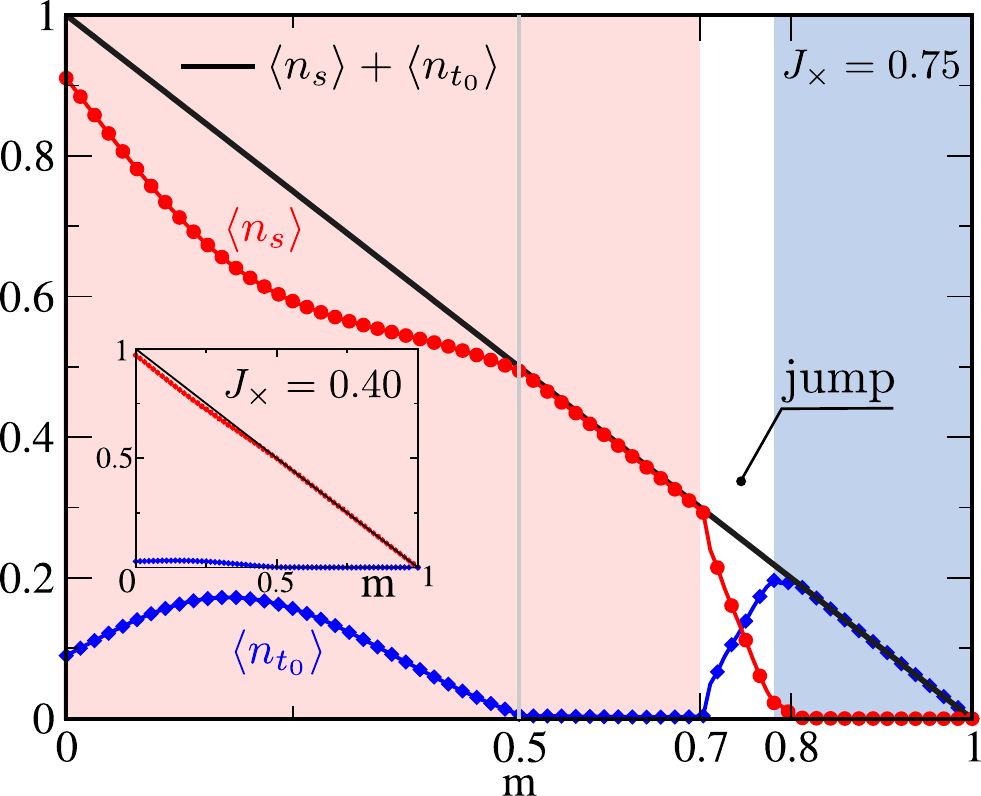}
    \caption{DMRG results for the average density of singlets $\braket{n_s}$ and triplets 
    $\braket{n_{t_0}}$ as a function of the magnetization $m$ for a system of size $L=128$, with $J_{\parallel}=0.55$ and $J_{\times}=0.75$. The magnetization states inside the magnetization jump are observed for $0.7\lesssim m\lesssim 0.78$. In the inset: the same parameters as in the main figure, except that $J_{\times}=0.40$.}
    \label{fig:singtrip0}
\end{figure}

We investigate the total probability of occurrence of dimer singlets and triplets through the 
two probability density parameters $\braket{n_s}$ and $\braket{n_{t_0}}$:
\begin{align}
 \braket{n_s}&=\frac{1}{L}\sum{\braket{n^s_l}},\nonumber\\
 \braket{n_{t_0}}&=\frac{1}{L}\sum{\braket{n^{t_0}_l}}.
\end{align}
In Fig. \ref{fig:singtrip0}, we show the behavior of
$\braket{n_s}$ and $\braket{n_{t_0}}$ for $J_\times=0.75$ and $J_\times=0.4$, in the inset, 
for $0.5<m<1.0$. In particular, we notice that the data cross the first-order transition line $h_f$ 
of the phase diagram (shown in Fig. \ref{fig:fase}) for $J_\times = 0.75$. For
$m_f<m<1$, with $m_f\approx 0.78$, we have $\braket{n_s}=0$ and $\braket{n_{t_0}}\neq 0$, while $\braket{n_s}\neq0$ and $\braket{n_{t_0}}=0$ for $0.5<m<0.70$. There is also a range of magnetization values, $\Delta m=m_f-m_i=0.78-0.70=0.08$, for which phase coexistence is observed in spatially separated regions, with $\braket{n_{t_0}}\neq 0$ and $\braket{n_{s}}\neq 0$. For $J_\times=0.4$, in the inset of Fig. \ref{fig:singtrip0}, the first-order
transition line is not crossed, see Fig. \ref{fig:fase}, and we have $\braket{n_{t_0}}= 0$ and $\braket{n_{s}}\neq 0$ for
$0.5<m<1$. For $m<0.5$, there is the coexistence of both components along
the total extension of the ladder, such that any dimer is in a coherent superposition of both components.

The quasi-long-range magnetic order in the high-field regime can be described by considering
$\braket{n_s}$, $\braket{n_{t_0}}$, and the transverse spin correlation functions.
We consider two types of transverse spin correlation functions: the first
between spins along the same leg $\Gamma_{11}(r)=\Gamma_{22}(r)$, and the other between spins
in different legs $\Gamma_{12}(r)=\Gamma_{21}(r)$. In both cases, to reduce boundary effects, we average ($\langle \cdots \rangle_{|m-l|=r}$) 
the transverse spin correlation $\langle S^+_{l,i}S^-_{m,j}+S^-_{l,i}S^+_{m,j}\rangle$ over all pairs of dimers $l$ and $m$ 
separated by the same distance $r$, such that: 
\begin{equation}
 \Gamma_{ij}(r)=\frac{1}{2}\langle\langle S^+_{l,i}S^-_{m,j}+S^-_{l,i}S^+_{m,j}\rangle\rangle_{|m-l|=r}.
 \label{eq:transvcorr}
\end{equation}

\begin{figure}
    \centering
    \includegraphics[width=0.48\textwidth]{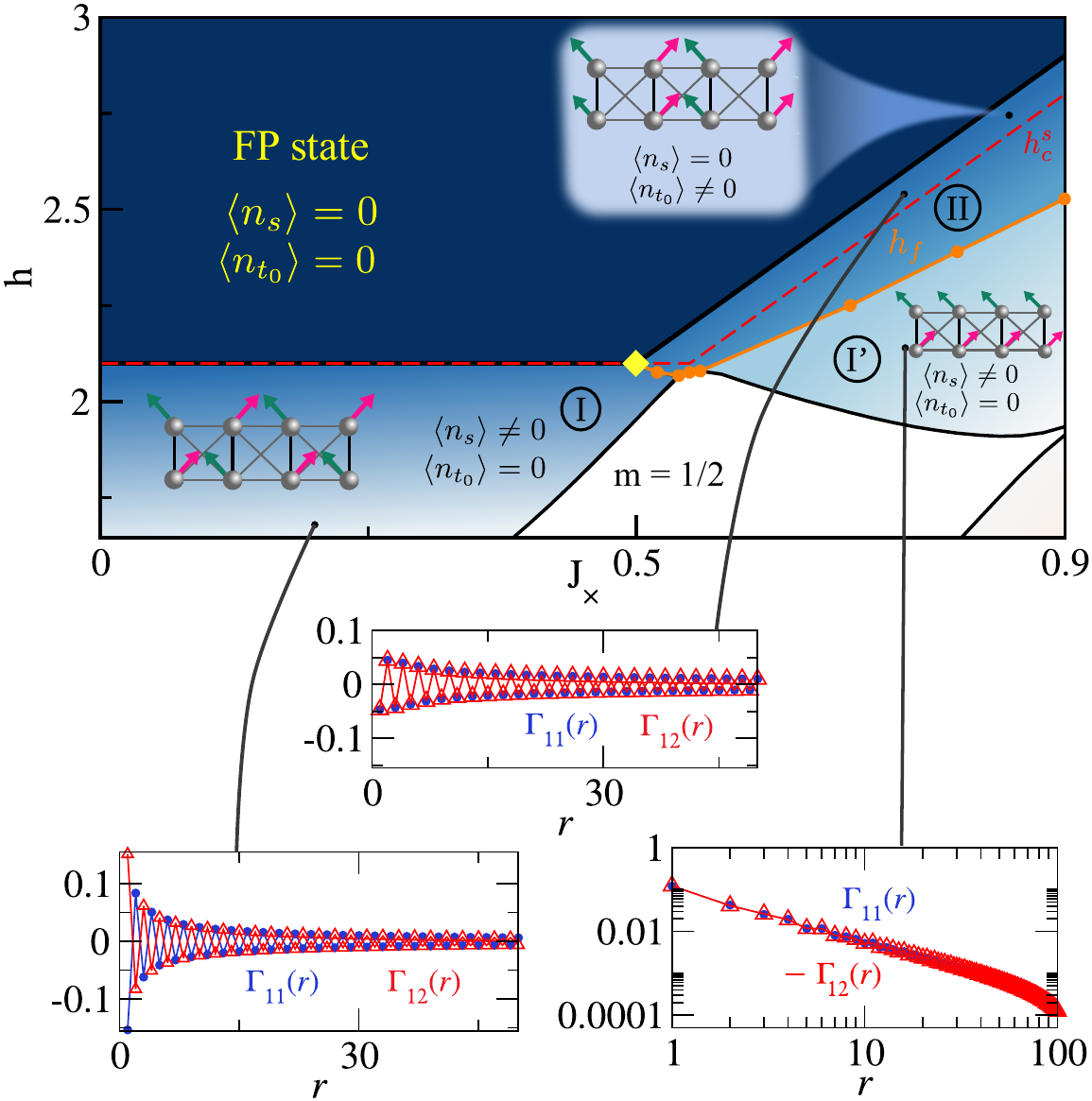}
    \caption{Short-range magnetic ordering near the fully polarized (FP) magnetization plateau as indicated 
    by the transverse spin correlation functions along one of the legs,  $\Gamma_{11}(r)$, and between spins 
    in distinct legs, $\Gamma_{12}(r)$, shown in the bottom panels. 
    In regions I and I$^\prime$, the average density of singlets $\braket{n_s}\neq 0$ and the average density of triplets $\ket{t_0}$ is $\braket{n_{t_0}}\approx 0$, 
    while in region II $\braket{n_s}\approx 0$ and $\braket{n_{t_0}}\neq 0$. We indicate the bicritical point ($\color{yellow}\blacklozenge$) at $J_\times=0.5$ and $h=1+2J_{\parallel}=2.1$. The dashed line $h^s_c$ is the critical line for the condensation of the singlet component in the noninteracting model, while $h_f$ is the thermodynamic limit first-order transition line from DMRG.}
    \label{fig:h_m1}
\end{figure}

We sketch in Fig. \ref{fig:h_m1} the short-range (power-law decaying) magnetic orders in the phases, 
named I, I$^\prime$, and II, which are competing in the high-field regime. 
The values of $\braket{n_s}$ and $\braket{n_{t_0}}$ in each phase are indicated, and we present
the behavior of $\Gamma_{11}(r)$ and $\Gamma_{12}(r)$ for representative points in these phases. 

First, we discuss the presence of a bicritical point \cite{chaikin2000principles,Fisher1974,Kosterlitz1976} at $J_{\times}=0.5$ and $h=2.1$. At this point, the two critical phases become identical to the disordered fully polarized phase. The bicritical point can thus be characterized by probability densities $\braket{n_s}$ and $\braket{n_{t_0}}$, as defined 
in the Eqs. \ref{eq:orderns} and \ref{eq:ordert0}, respectively. The disordered phase, the fully polarized magnetization, is
gapped with $\braket{n_s}=0$ and $\braket{n_{t_0}}=0$. 
For $J_{\times}<0.5$, as the field is reduced, the singlet component
condenses, and the average probability density $\braket{n_{s}}\neq 0$ in phase I, with $\braket{n_{t_0}}=0$.
For $J_{\times}>0.5$, as the field is reduced, the component $\ket{t_0}$ condenses, and thus we find an average probability density $\braket{n_{t_0}}\neq0$ in phase II,  
with $\braket{n_{s}}=0$. In fact, in phase II, the frustrated ladder is effectively described by a spin-1 chain
in a magnetic field. We mention that in the symmetrical case, $J_\times = J_\parallel$, the effective description of the frustrated ladder by spin-1 chains was revealed for $h=0$ in Ref. \cite{Gelfand1991}, and in Ref. \cite{Honecker_2000} for $h\neq 0$. Furthermore, the equivalence between the low-energy behavior of frustrated ladders and spin-1 chains was also observed for other frustration patterns and zero magnetic field \cite{White1996}.

The transition from phase II, with $\braket{n_{t_0}}\neq0$ and $\braket{n_{s}}= 0$, 
to phase I or phase I$^\prime$, both with $\braket{n_{t_0}}=0$ and $\braket{n_{s}}\neq0$, is of first order and is indicated by the transition line $h_f$. Furthermore, near $J_{\times}=J_{\parallel}=0.55$, the transition from the plateau $m=1/2$ to phase II is also of the first-order kind.  

We notice that the line for the condensation of the
singlet component $h_c^s$, Eqs. (\ref{eq:hs1}) and (\ref{eq:hs2}), from the free hard-core bosons model in (\ref{eq:hamboson}), also shown in Fig. \ref{fig:h_m1}, has a trend that resembles the line $h_f$ from the DMRG data. In particular, for $J_\times<J_\parallel$, $h_c^s=1+2J_\parallel=2.1$ 
is constant, and $h_c^s=1+2J_\times=2.1$ for $J_\times>J_\parallel$, having an upward slope. Compared with it, the DMRG results for $h_f$ show that the interactions have introduced an average downward slope for the
line in the region $J_\times>J_\parallel$, and also for the horizontal line in the range $0.5<J_\times<J_\parallel$.

The phases I, I$^\prime$, and II are gapless phases in which dimers are predominantly found in a coherent superposition
of triplets $\ket{\uparrow\uparrow}$ and singlets (phases I and I$^\prime$) or triplets $\ket{t_0}$ (phase II). Thus, the dimer
spins present the classical orientations depicted in Fig. \ref{fig:freebands}(d). The canted orientation of the dimer
spins and the competition between $J_{\times}$ and $J_{\parallel}$ explains the quasi-long-range magnetic order of phases I and I$^\prime$.
For $J_{\times}<J_{\parallel}$ there is a quasi-long-range transverse antiferromagnetic order between spins in the
same leg; while for $J_{\times}>J_{\parallel}$ this order is observed between spins of distinct legs. 
Phases I and I$^\prime$ meet at the gapped $m=1/2$ magnetization, fully frustrated 
point $J_{\times}=J_{\parallel}=0.55$, for which the singlet and
triplet $\ket{\uparrow\uparrow}$ components are localized, as will be discussed in the next section. In contrast to phases
I and I$^\prime$, in phase II, there is no competition between $J_{\times}$ and $J_{\parallel}$, since,
in this case, the coherent superposition between the triplet components $\ket{t_0}$ and $\ket{\uparrow\uparrow}$ satisfies
both couplings. In this case, the transverse spin correlation functions present an antiferromagnetic quasi-long-range order
between spins along the legs and between spins at distinct legs.

\subsection{phase separation}
\begin{figure}
    \centering
    \includegraphics[width=0.48\textwidth]{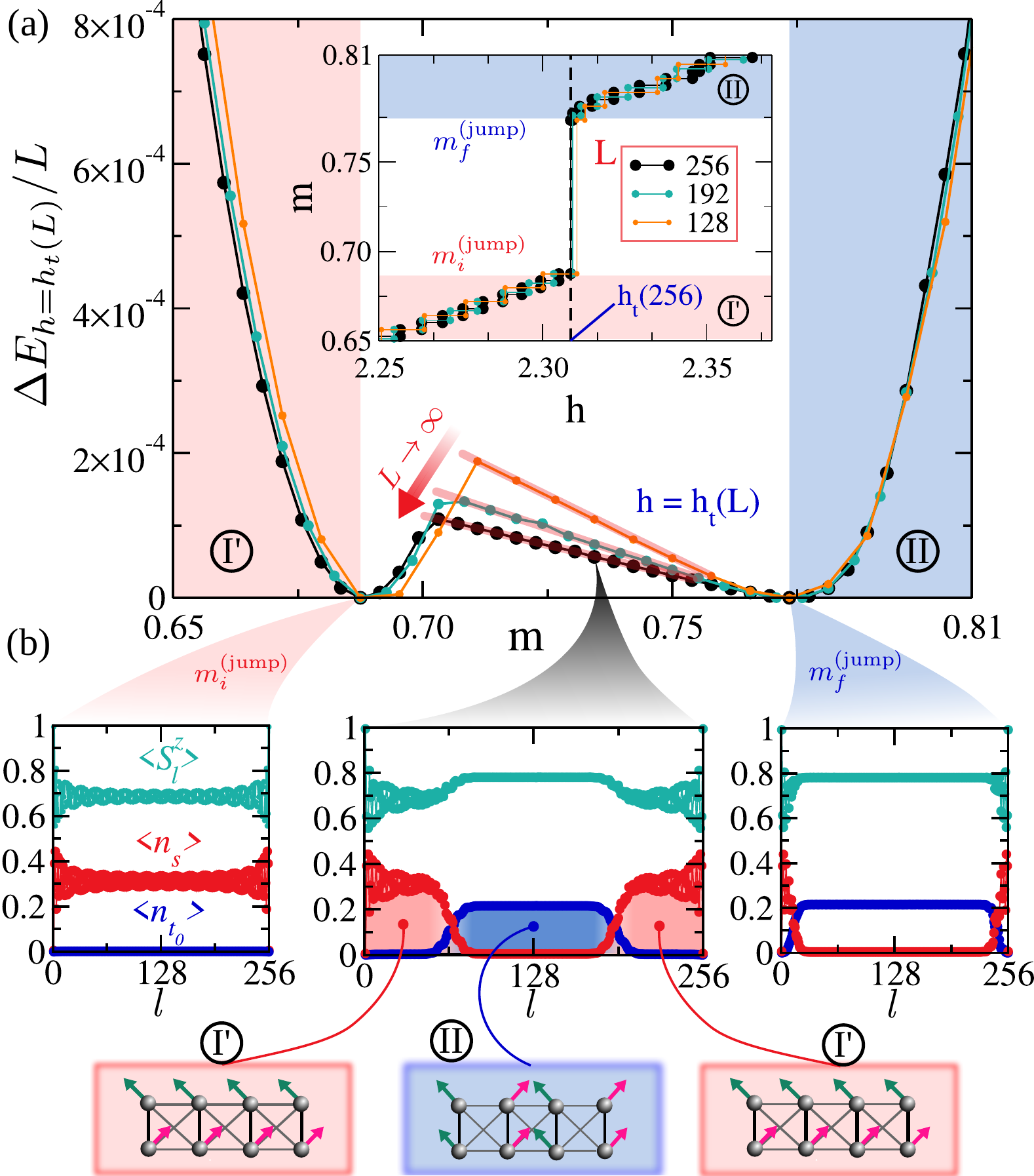}
    \caption{(a) The Difference between the energy per dimer [$E_{h=h_t(L)}/L$] and its minimum: 
    $\Delta E_{h=h_t(L)}/L=(E_{h=h_t(L)}-E_{h=h_t(L),min})/L$, as a function of the magnetization $m$ near the magnetization jump for $J_\times=0.74$, $J_\parallel=0.55$, and magnetic field $h=h_t(L)$, where $h_t(L)$ is the value of $h$ at which the jump is observed in systems of size $L=128,~192,~256$. The inset displays $m$ as a function of $h$ for the magnetization range shown in the main figure. We indicate the lower, $m_i^{(\text{jump})}$, and the upper, $m_f^{(\text{jump})}$, values of the magnetization that limit the magnetization jump. 
    In (b) we exhibit the magnetization distribution, $\braket{S^z_l}$, and average probability densities of singlet, $\braket{n_s}$, and triplet $\ket{t_0}$, $\braket{n_{t_0}}$, states along a chain of size $L=256$ for $m=m_i^{(\text{jump})}$, $m_f^{(\text{jump})}$, and an intermediate value of $m$ ($=188/256$), as indicated in (a). In the central panel, we illustrate the phases in each region of the chain (see also Fig. \ref{fig:h_m1}).}
    \label{fig:enps}
\end{figure}

Along the first-order transition line $h=h_f(J_{\times})$ shown in Figs. \ref{fig:fase} and \ref{fig:h_m1}, 
we expect phase coexistence between phase II and phases I or I$^\prime$. 
Around $h=h_f(J_{\times})$ the competition occurs between a singlet rich phase (I or I$^\prime$) 
and a triplet $\ket{t_0}$ rich phase (II). In the thermodynamic limit, precisely at $h=h_f$, the energy
curve as a function of $m$ has a flat region, with degenerate magnetization states in the range $m_i^{(\text{jump})}<m<m_f^{(\text{jump})}$, 
where $m_i^{(\text{jump})}$ and $m_f^{(\text{jump})}$ are the extreme magnetizations of the jump. 
At $m=m_i^{(\text{jump})}$ (phases I or $I^\prime$), we should have $\braket{n_s}\neq 0$ and $\braket{n_{t_0}}=0$, while
at $m=m_f^{(\text{jump})}$ (phase II), $\braket{n_s}= 0$ and $\braket{n_{t_0}}\neq 0$, see Figs. \ref{fig:h_m1} and \ref{fig:tudo}.
In the flat portion of the energy curve, $m_i^{(\text{jump})}<m<m_f^{(\text{jump})}$, the states exhibit spatial
separation of phases I or I$^\prime$ (singlet rich) and II (triplet rich).
Thus, to complete the whole physical picture associated with the first-order transition line and the bicritical point, we investigate the distribution of $\braket{n_s}$ and $\braket{n_{t_0}}$ along the ladder near the transition.

We show the energy per dimer $E_{h=h_t(L)}/L$ as a function of $m$ at the
respective value of the transition field $h=h_t(L)$ in Fig. \ref{fig:enps}(a). The curves are translated by the respective value of the energy
minimum, $E_{h=h_t(L),min}/L$. In the inset of this figure, we have put the finite-size magnetization curves as a reference. In particular, we notice the presence of
two minima separated from each other by a region of unstable states \textit{in finite-size systems}. 
\begin{figure}
    \centering
    \includegraphics[width=0.48\textwidth]{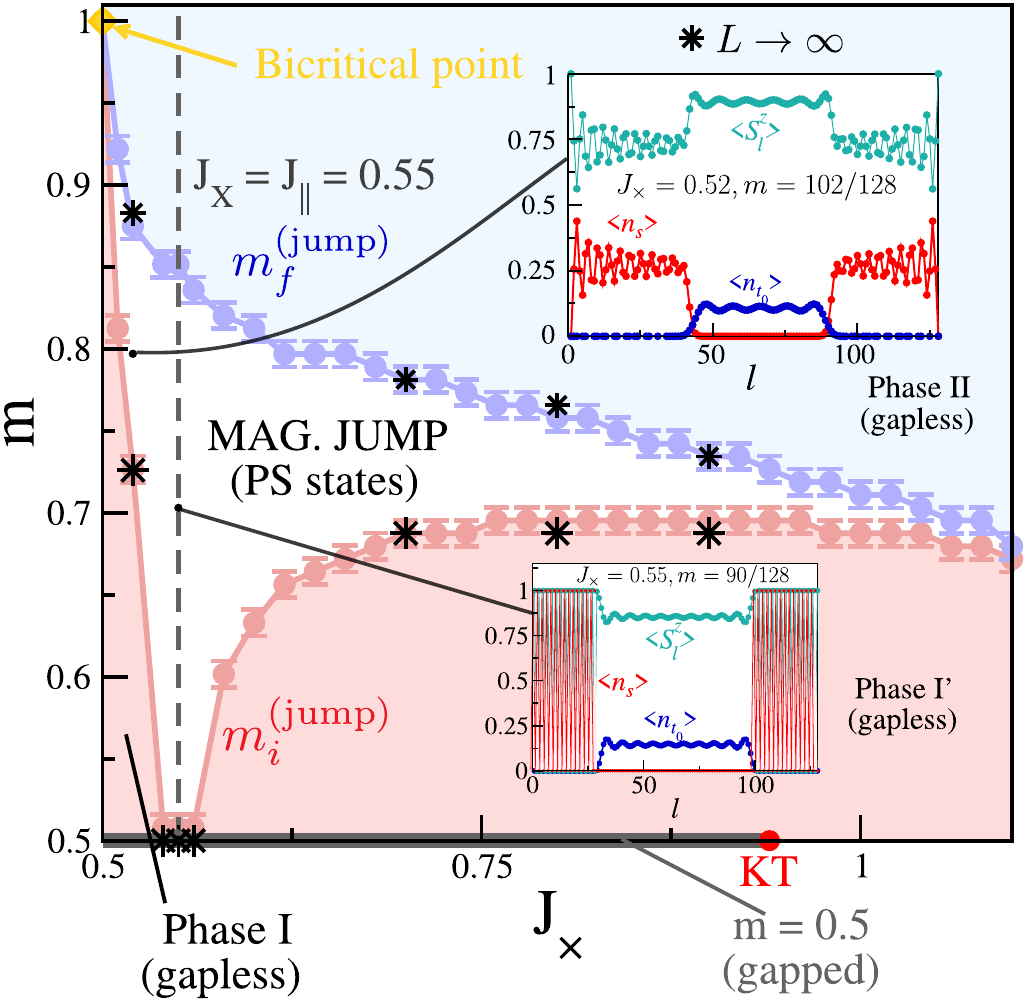}
    \caption{DMRG results for the lower, $m_i^{(\text{jump})}$, and upper, $m_f^{(\text{jump})}$, magnetization values 
     bounding the magnetization jump as a function of $J_\times$ for $J_\parallel=0.55$, ($\bullet$) $L=128$, and ($\ast$) $L\rightarrow \infty$. Error bars are defined as the minimum $\Delta m$ value for the system size: $\Delta m =1/128$. Phases I, I$^\prime$, and II are those sketched in Fig. \ref{fig:h_m1}. The bicritical point at $J_{\times}=0.5$ and $m=1$ ($h=2.1$ in Fig. \ref{fig:h_m1}), as well as the Kosterlitz-Thouless (KT) point, are also indicated. Inset: probability density of singlets ($\braket{n_s}$) and triplets $\ket{t_0}$ ($\braket{n_{t_0}}$) along the chain for $J_\times=$ 0.52 and 0.55, and $m=102/128$ and 90/128, respectively.}
    \label{fig:ps}
\end{figure}

In Fig. \ref{fig:enps}(b), we present the magnetization, singlet,
and triplet $\ket{t_0}$ distributions along the chain. 
Data show that at minimum $m=m_i^{(\text{jump})}$, the ladder is in
the singlet-rich phase I$^\prime$, while at minimum $m=m_f^{(\text{jump})}$, its bulk
is in the triplet-rich phase II. Furthermore, as shown in the central panel of Fig. \ref{fig:enps}(b), the states in the straight line portion of the energy curves exhibit phase separation. 

A finite-size scaling analysis of the angular coefficient of the straight line portion in the energy curves of Fig. \ref{fig:enps}(a), the region highlighted, shows that the thermodynamic-limit straight line is horizontal, joining the two single-phase states, and the energy density as a function of $m$ is an equilibrium curve. Furthermore, the size dependence of the energy curves
in the linear regime implies that the energy density of phase I$^\prime$ differs from the energy density of phase II by a term $a/L$, where $a$ is a constant. Since this difference is in the size $L$ of the whole system, we attribute it to the interface between the two coexisting states. 

In Fig. \ref{fig:ps}, we present the phase diagram $m$ versus $J_\times$ for $0.5\leq m \leq 1$ 
and $0.5\leq J_\times\leq 1.12$ for a ladder with 128 dimers. We are interested
in observing the magnetization region bounded by $m_i^{(\text{jump})}$ and $m_f^{(\text{jump})}$ 
as a function of $J_{\times}$, particularly to show that $m_i^{(\text{jump})}\rightarrow m_f^{(\text{jump})} \rightarrow 1$ as $J_\times\rightarrow0.5$, the quantum bicritical point. 
We also show singlet and triplet distributions of $\ket{t_0}$, $\braket{n^s_l}$ and $\braket{n^{t_0}_{l}}$, along a 128 dimer ladder for $(J_\times=0.52,~m\approx 0.8)$ and $(J_\times=0.50,~m\approx 0.7)$.
The region of phase separation is limited by the bicritical point, $(J_\times=0.50,~m=1)$, on the left,
and we could identify not a jump but a discontinuity in the slope of the magnetization curve at $J_{\times}\gtrsim 1.1$ on the right. For magnetizations $m_i^{(\text{jump})}<m<m_f^{(\text{jump})}$, the gapless phase II coexists with the gapless phase I for $0.5<J_{\times}<0.55=J_{\parallel}$, and with the gapless phase I$^{\prime}$ for $J_{\times}>0.55$. On the other hand, phase II coexists with the gapped $m=1/2$ state at $J_{\times}=J_{\parallel}$ and $m_i^{(\text{jump})}<m<m_f^{(\text{jump})}$. 

In particular, we discuss a technical difficulty in obtaining the distributions for $J_{\times}=J_{\parallel}=0.55$. 
In this case, the gapless phase II coexists with the gapped $m=0.5$  
state, and the renormalization procedure becomes more complex. 
For $J_{\times}=J_{\parallel}$, the Hamiltonian is invariant under the exchange of spin variables $\mathbf{S}_{l,1}$ and $\mathbf{S}_{l,2}$ in the same dimer. Therefore, each dimer has constant parity. 
Since the Hamiltonian cannot connect different parity sectors, the parity of the dimers
is fixed in the growth stage of the renormalization procedure and cannot change in the sweeping stage of the DMRG.
Also, we notice that the singlet band (\ref{eq:dispersion2}) becomes flat at $J_{\times}=J_{\parallel}$, 
implying that the interaction controls the physics at this point.
Due to these issues, if we run the DMRG procedure targeting only the desired value of $m$,
the algorithm converges to a higher energy state 
presenting a phase-separated state with clusters of variable size
of phase II separated by clusters of the $m=1/2$ state. 
However, the lowest energy state shown in Fig. \ref{fig:ps} has only two domains and can be reached by running
the DMRG procedure twice. First, we target the final value
of $m$ and, at the end of the run, we estimate the total size $l_0$ of the $m=1/2$ phase as the sum of the sizes
of the clusters of this phase.
In the second run, we target the sector $m=1/2$ in the growth stage until the chain has a size equal to $l_0$, and after that we target
the desired value of $m$. At this time, the wavefunction converges to the
phase-separated state shown in Fig. \ref{fig:ps}, which has lower energy than the one obtained in the first run.

\section{$m=1/2$ magnetization plateau and the Kosterlitz-Thouless transition points}
\label{sec:kttransitions}
\begin{figure}
    \centering
    \includegraphics[width=0.28\textwidth]{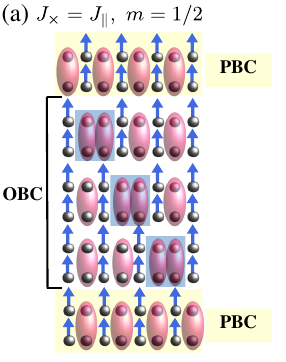}
    \includegraphics[width=0.385\textwidth]{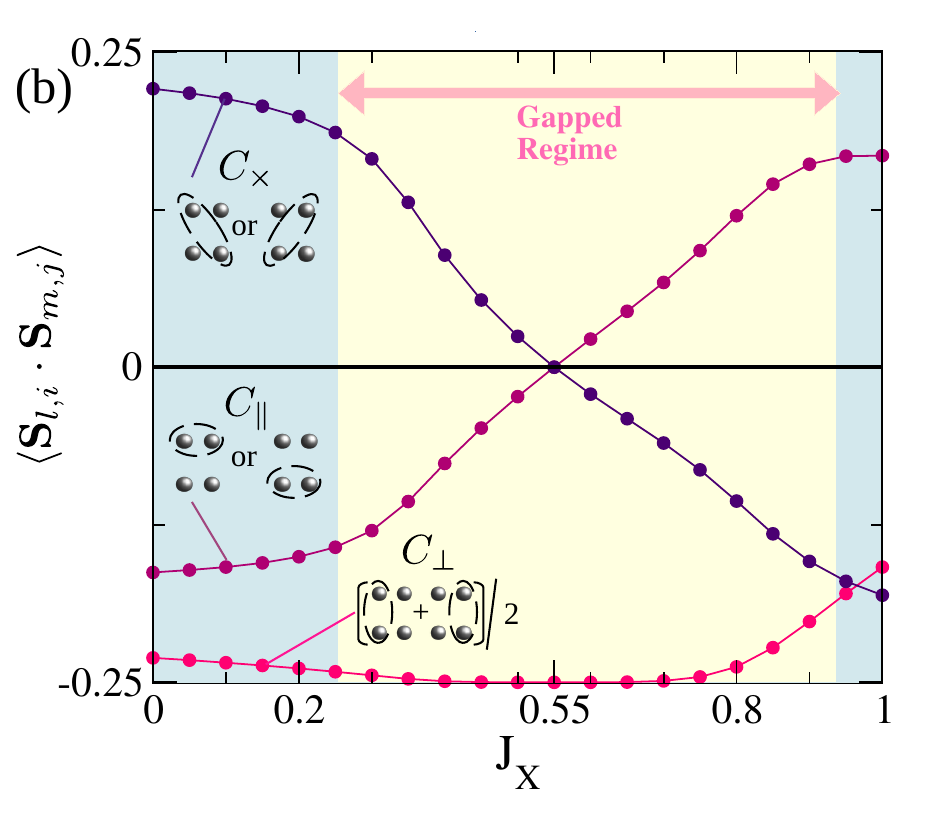}
    \caption{(a) Degenerate states with $m=1/2$ for $J_{\times}=J_{\parallel}$ in a chain with open boundary conditions (OBC) or periodic boundary conditions (PBC). (b) Exact diagonalization results for a chain with $L=14$ and $J_{\parallel}=0.55$ with PBC. Correlation functions $\langle \mathbf{S}_{l,i}\cdot\mathbf{S}_{m,j}\rangle$ between the spins at the same dimer ($l=m$), $C_{\perp}$, and neighbor dimers ($l=m+1$), $C_{\times}$ or $C_{\parallel}$, as illustrated by the dashed ellipses. The estimated gapped regime in the thermodynamic limit is indicated.}
    \label{fig:spinon}
\end{figure}

The magnetization plateau state $m=1/2$ can be better qualitatively understood from
the $J_{\times}=J_{\parallel}$ line \cite{Honecker_2000,Fouet2006}. Because all dimers have a fixed parity, they cannot be found in a coherent superposition of singlet and triplet states. The ground-state 
wavefunction has a singlet alternating with a triplet $\ket{\uparrow\uparrow}$ along the
lattice due to the repulsion energy between two triplets
if they are in neighbor dimers. In a finite chain with periodic boundary conditions (PBC) \cite{Fouet2006}, 
there are two degenerate states, as illustrated in Fig. \ref{fig:spinon}(a). However, for
a finite chain with open boundary conditions (OBC), the lowest energy states have triplets $\ket{\uparrow\uparrow}$ at the edges of the chain due to an edge term in the Hamiltonian \cite{Fouet2006}. 
As also illustrated in Fig. \ref{fig:spinon}(a), these triplets cause the presence of two neighboring dimers in a singlet state
along the chain. 
This pair of singlets is the edges of two domains, each of them having the alternation of one of the PBC ground states.
Even in that case, $J_{\times}=J_{\parallel}$, the ground state exhibits extensive degeneracy, since the
energy does not change as we change the position of the pair of singlets. However, this degeneracy is lifted for $J_{\times}\neq J_{\parallel}$.
This domain wall or spinon carries a spin-1/2 and was studied in detail in Ref. \cite{Fouet2006}.
At the magnetic field $h=h_0$, see Fig. \ref{fig:tudo}, one of the singlets in the domain wall pair
changes to a triplet $\ket{\uparrow\uparrow}$, and the spin of the spinon changes from -1/2 to +1/2.

We use exact diagonalization for a chain with
$L=14$ dimers and PBC to show in Fig. \ref{fig:spinon}(b) the correlations between the neighbor spins connected by each of the couplings: $J_{\perp}\equiv 1$, $J_{\times}$, and $J_{\parallel}=0.55$, which we call $C_{\perp}$, $C_{\times}$, and $C_{\parallel}$, respectively, as functions of $J_{\times}$. In particular, due to the twofold degeneracy for $J_{\times}=J_{\parallel}$, Fig. \ref{fig:spinon}(a), $C_{\perp}$ is defined as the average of this correlation for two neighbor dimers. For $J_{\times}=J_{\parallel}$, the singlet components are fixed, so $C_{\perp}=-0.25$, since the dimer singlets are sided by two dimer triplets; and $C_{\times}=0=C_{\parallel}$. 
For $J_{\times}<J_{\parallel}$, $C_{\times}>0$ and $C_{\parallel}<0$, thus there is a higher probability that the singlet component will fluctuate from the perpendicular to the parallel direction; while in the case
of $J_{\times}>J_{\parallel}$ the roles of $C_{\parallel}$ and $C_{\times}$ are exchanged in relation to the first
case. However, we notice that the values of $C_{\parallel}$ and $C_{\times}$ are not exactly exchanged about the $J_{\times}=J_{\parallel}$ point, since $J_{\times}\rightarrow 1= J_{\perp}$ on the right side of the figure. In particular, 
$C_{\times}$ is lower than $C_{\perp}$ for $J_{\times}\gtrsim 0.97$. 

In a short-range region of size $\sim \xi$, the spin correlations are similar to those of
phase I, illustrated in Fig. \ref{fig:h_m1}, for $J_{\times}<J_{\parallel}$; and of phase I$^\prime$ for $J_{\times}>J_{\parallel}$. At the KT transition points, $\xi\rightarrow\infty$ and
the correlations will exhibit a power-law behavior, as shown in Fig. \ref{fig:h_m1}. In the following, we estimate
the critical points of the KT transitions.

\subsection{Kosterlitz-Thouless transitions}

In the Kosterlitz-Thouless transitions, the value of $m$ is fixed at 1/2, and
the gap $\Delta h$ closes following an essential
singularity form.
The asymptotic behavior of $\Gamma_{ij}(r)$
on the gapless side of the transition is given by Eq. (\ref{eq:ll-corr})
with $K=1/2$, since the translation symmetry is broken and we have one boson for every two dimers
\cite{Cazalilla_2011} of the chain. 
A confident numerical technique to estimate the value of $J_c$ from finite-size 
systems with open boundaries uses the exact expected value of $K$ as $J\rightarrow J_c$. 
Within this methodology \cite{Kuhner,MontenegroFilho2020}, the thermodynamic limit value of $K$ 
is estimated through an unbiased approach, as detailed below.

\begin{figure}
    \centering
    \includegraphics[width=0.48\textwidth]{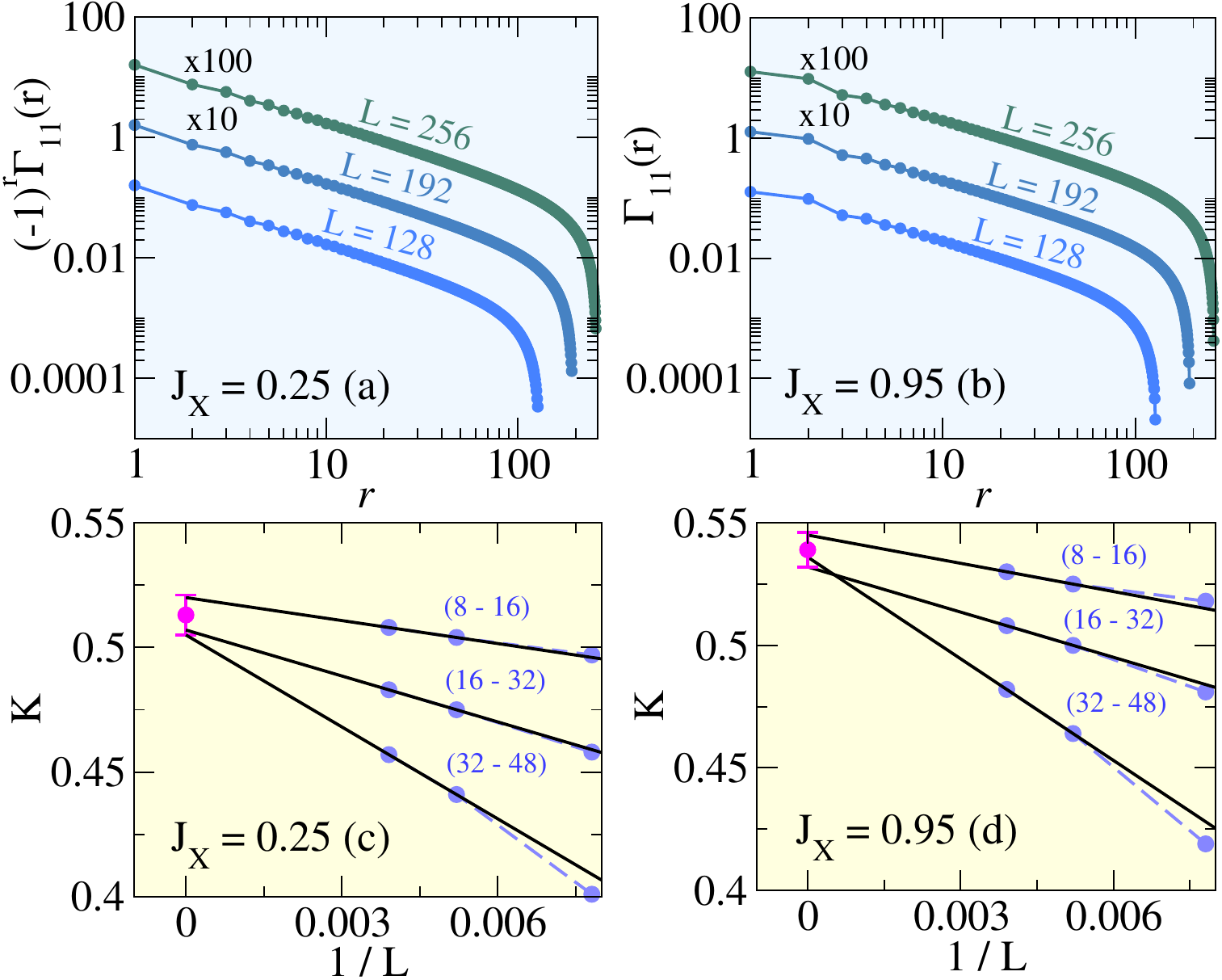}
    \caption{Transverse spin correlation functions $\Gamma_{11}(r)$, where $r$ is the distance along the chain, and Luttinger liquid exponent $K$ for $m=1/2$ and $J_{\parallel}=0.55$. In (a), we present $(-1)^r\Gamma_{11}(r)$ for $J_{\times}=0.25$; while in (b) we show $\Gamma_{11}(r)$ for $J_{\times}=0.95$, 
    in both cases for three system sizes: $L=128,~192,~\text{and}~256$. In (c) and (d): The Luttinger liquid exponent as a function of $1/L$. $K$ is calculated by fitting the correlations in (a) and (b) to the form $1/r^{1/2K}$ along the following intervals of $r$: $8\leq r\leq 16$, $16\leq r\leq 32$, and $32\leq r\leq 48$, for (c) $J_{\times}=0.25$ and (d) $J_{\times}=0.95$. 
    Extrapolating the values of $K$ for each interval of $r$ using a straight line through the two larger sizes, we obtain the minimum, $K_{min}$, and maximum, $K_{max}$, values of $K$ for a given $J_{\times}$. 
    The thermodynamic limit value of $K$ is defined as $(K_{min}+K_{max})/2$, and the error in $K$ as $(K_{max}-K_{min})/2$.}
    \label{fig:corre}
\end{figure}

The transverse spin correlation functions of finite-size systems with open boundaries
are calculated through Eq. (\ref{eq:transvcorr}).  In Figs. \ref{fig:corre} (a) and (b), we
present the typical behavior of $\Gamma_{11}(r)$ for two values of $J_{\times}$ and $m=1/2$, see also Fig. \ref{fig:fase}. 
We must fit the data to the asymptotic form of the correlation, Eq. \ref{eq:ll-corr}, to obtain $K$. 
However, in a system with open boundaries, the value of $K$ thus obtained strongly depends
on the interval of $r$ chosen to make the fit. To overcome this problem \cite{Kuhner}, we
arbitrarily define some intervals of $r$ and extrapolate the finite-size values of $K$ to the
thermodynamic limit for each interval. In Figs. \ref{fig:corre}(c) and (d), we present this extrapolation for
the same values of $J_{\times}$ used to calculate the correlations shown in Figs. \ref{fig:corre}(a) and (b). 
In the extrapolation, we use a linear scale function to fit the points of the two largest system sizes. 
Notice that the extrapolated values of $K$ for the three $r$ intervals are very similar and have little dispersion. The minimum and maximum
values of $K$, $K_{min}$, and $K_{max}$, respectively, of the three intervals are used
to define our thermodynamic limit estimate as $K=(K_{max}+K_{min})/2$, with error $(K_{max}-K_{min})/2$. 

\begin{figure}
\includegraphics[width=0.48\textwidth]{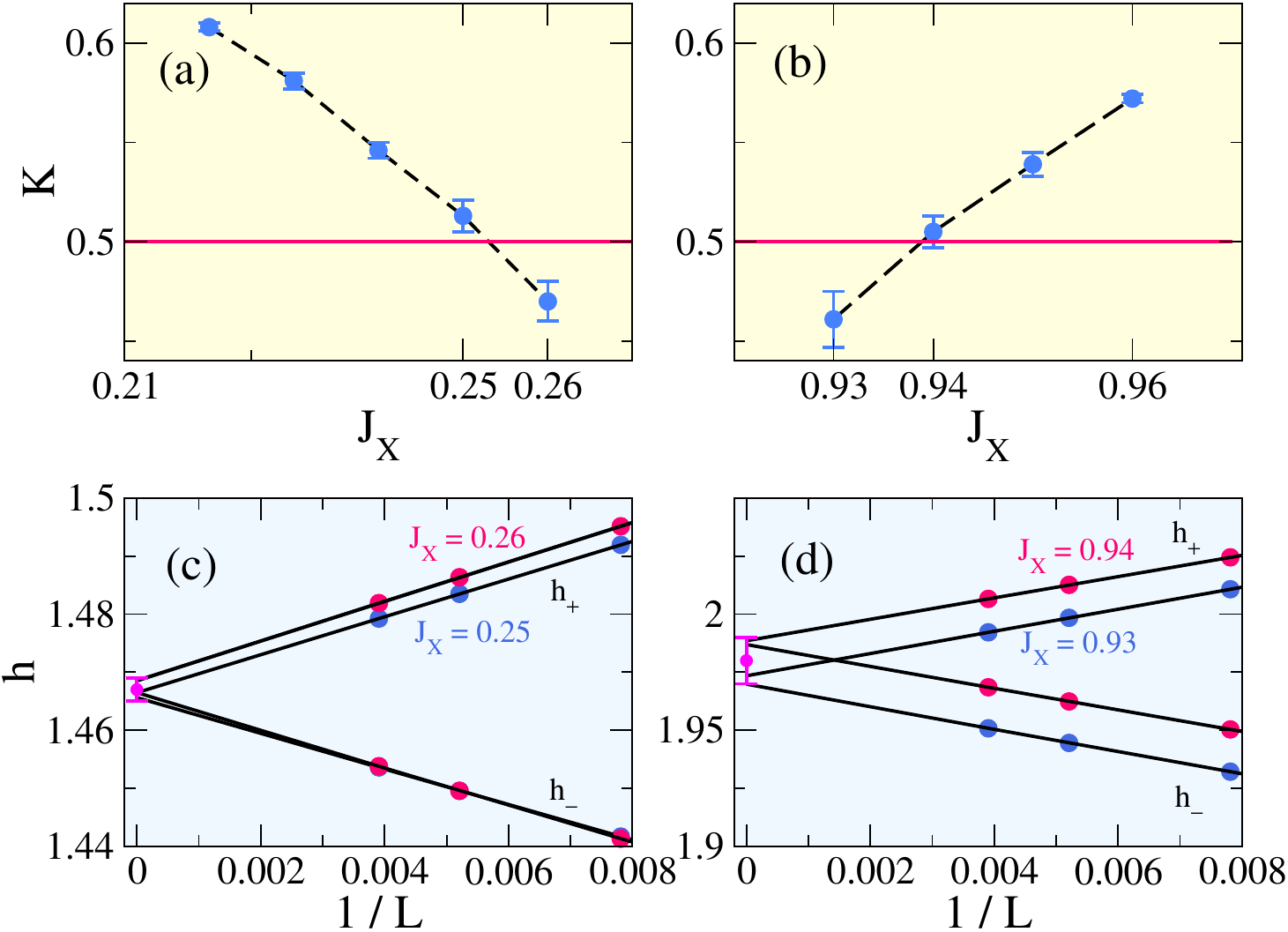}
\caption{(a) and (b): Thermodynamic limit estimate for the Luttinger parameter $K$ as a function of $J_{\times}$ 
    around the Kosterlitz-Thouless transition points. We estimate the critical points as $J_{\times, \text{KT}_1}=0.255\pm0.005$ 
    and $J_{\times, \text{KT}_2}=0.935\pm0.005$. (c) and (d): extrapolation of the critical values of the magnetic field $h$, $h_-$, and $h_+$, for $m=1/2$ to the thermodynamic limit. We estimate $h_{\text{KT}_1}=1.467\pm0.002$ and $h_{\text{KT}_2}=1.98\pm0.01$.}
    \label{fig:k_crit}
\end{figure}

In Fig. \ref{fig:k_crit} (a) we show $K$ as a function of $J_{\times}$ near the KT transition
in the region $J_{\times}<J_{\parallel}$; while in Fig. \ref{fig:k_crit} (b) we show $K$ near the
transition in the region $J_{\times}>J_{\parallel}$, see also Fig. \ref{fig:fase}. 
Through these data, we estimate the values of $J_{\times}$ for the two critical points as
$J_{\times, \text{KT}_1}=0.255\pm0.005$ and $J_{\times, \text{KT}_2}=0.935\pm0.005$.
To estimate the critical fields, we extrapolate to the thermodynamic limit
the extreme fields, $h_{-}$ and $h_{+}$, of the magnetization steps
for $m=1/2$ in finite-size systems. We take $h_{-}$ and $h_{+}$ for $J_{\times}$ 
just before and just after $J_c$, considering the data shown in Figs. \ref{fig:k_crit} (a) and (b).
We estimate the critical magnetic fields as $h_{\text{KT}_1}=1.467\pm0.002$ and $h_{\text{KT}_2}=1.98\pm0.01$ 
from the finite-size analysis shown in Figs. \ref{fig:k_crit} (c) and (d).

As was mentioned earlier, in the regime $J_{\perp}\gg (J_{\parallel}, J_{\times})$ and $h>0$, 
in first-order approximation \cite{Mila_1998, Fouet2006}, mappings onto the one-dimensional XXZ Heisenberg or spinless interacting
fermion models are possible. Within these models, the Bethe ansatz solution allows to locate precisely \cite{Mila_1998} the critical KT point at $J_{\times}=J_{\parallel}/3$, for $J_{\times}<J_{\parallel}$. In our case, $J_{\parallel}=0.55$, and the
value of the critical point is $J_{\times}=0.18$ in the approximate approach, departing from our prediction, 
$J_{\times, \text{KT}_1}=0.255\pm0.005$, by $\sim J_{\times, \text{KT}_1}^2$. 
For the second critical point, $J_{\times, \text{KT}_2}=0.935\pm0.005$,
the first-order approximation is not reliable since $J_{\times}$ and $J_{\parallel}$ have the same order of $J_{\perp}$.

\section{Other values of $J_\parallel$, and changing $J_\parallel$ with $J_\times$ constant}
\label{sec:othervalues}
\begin{figure*}
    \centering
    \includegraphics[width=\textwidth]{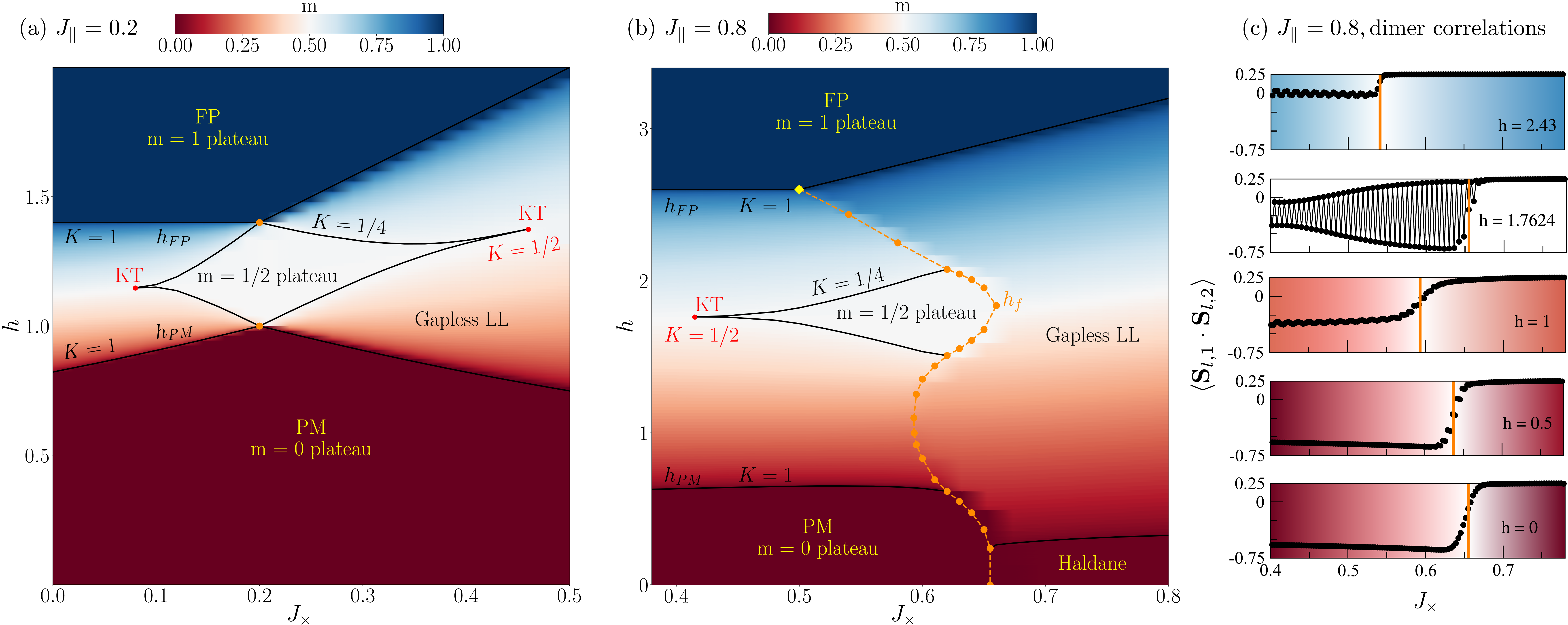}
    \caption{DMRG results for the magnetic field $h$ versus frustration $J_\times$ phase diagram for 
    (a) $J_\parallel=0.2$ and (b) $J_\parallel=0.8$. The thermodynamic limit transition lines are estimated from finite-size scale analysis of magnetization ($m$) as a function of $h$, while the color code is the value of $m$ for a system of size $L=128$. In the phase diagrams, we highlight the fully polarized (FP), gapped paramagnetic (PM) phase, gapless Luttinger liquid phases, the value of the Luttinger liquid exponent ($K$) in the incommensurate transitions and the Kosterlitz-Thouless (KT) transition points. The dashed line in (b) is a first-order transition line, and the quantum bicritical point is indicated with a ($\color{yellow}\blacklozenge$). We indicate the region at which the gapped singlet Haldane phase is observed. (c) We show $J_\times$-scans, for fixed $h$, of the intradimer correlation $\langle\mathbf{S}_{l,1}\cdot\mathbf{S}_{l,2}\rangle$. The vertical orange lines mark the first-order transition point in the thermodynamic limit, as shown in (b).
    }
    \label{fig:otherjp}
\end{figure*}

This section discusses the phase diagrams for $J_\parallel=0.2$ and $J_\parallel=0.8$, shown in Fig. \ref{fig:otherjp}, to grasp the stability of the features observed for $J_\parallel=0.55$. We also analyze the phase diagrams for fixed $J_\times$, varying $J_\parallel$, and the KT transition points in the plane $J_\times$ versus $J_\parallel$. 

As expected in Fig. \ref{fig:freehfp}, the bicritical point and the first-order transition line are absent in the phase diagram
for $J_\parallel=0.2$, Fig. \ref{fig:otherjp}(a), since the minimum energy of the triplet
and singlet bands does not cross for $J_\parallel<0.5$. Furthermore, from Fig. \ref{fig:freehfp}, we can state that
the bicritical point is located at $J_\times=0.5$ and $h=1+2J_\parallel$ for $J_\parallel>0.5$, as exemplified by
the phase diagram in Fig. \ref{fig:otherjp}(b), for $J_\parallel=0.8$.

For $J_{\parallel}=0.2$, there are only two first-order transition points, 
both at $J_\times=J_\parallel=0.2$: one that separates the plateaus $m=0$ and $m=1/2$, and
the other between the plateaus $m=1/2$ and $m=1$. However, as in the case of $J_\parallel=0.55$,  the closing of the plateau $m=1/2$ on the two sides of the phase diagram, $J_\times < J_\parallel$ and $J_\times > J_\parallel$, follows a KT transition for $J_\parallel=0.2$. 
The DMRG data show that the ground state is predominantly composed of a coherent superposition of the triplet component $\ket{\uparrow\uparrow}_l$ and $\ket{s_l}$, as in the cases
$J_\times=0.4$ and $J_\parallel=0.55$ shown in Fig. \ref{fig:singtrip0}. 

The first-order transition points at $(J_\times=J_\parallel=0.2,h=1)$ and $(J_\times=J_\parallel=0.2,h=h_{FP})$ are in the meeting of four second-order transition lines.
At these points, there is the coexistence of the two disordered phases: $m=0$, $m=1/2$ in the first case and $m=1/2$, $m=1$ in the second.  
At these points, the critical magnetization states of the Luttinger liquid phases: $0<m<1/2$ for $h=1$
and $1/2<m<1$ for $h=h_{FP}$, become equal to the disordered $m=0$ or $m=1/2$ phases and to the disordered $m=1/2$ or $m=1$ phases, respectively. We mention that a similar point is also observed at $(J_\times=J_\parallel=0.55,h=1)$ in the phase diagram of Fig. \ref{fig:fase}. 

In Landau theory for thermal phase transitions  \cite{Fisher1974,chaikin2000principles}, the \textit{tetracritical} point is also found at the meeting of four second-order transition lines. In that case, there are three ordered phases with two distinct orders. In one of the phases, the ``intermediate phase'' \cite{Fisher1974,chaikin2000principles}, the two order parameters are finite; while in the other two phases, one or the other parameter
is different from 0. The point $(J_\times=J_\parallel=0.2,h=1)$ in the phase diagram of Fig. \ref{fig:otherjp}(a) could be a tetracritical point if one of the two phases $m=0$ or $m=1/2$ was critical or ordered, instead of disordered. A similar discussion can be had for the two similar first-order points mentioned: in $(J_\times=J_\parallel=0.2,h=1)$ and $(J_\times=J_\parallel=0.55,h=1)$.  

The phase diagram is richer for $J_{\parallel}=0.8$, Fig. \ref{fig:otherjp}(b). In particular, we notice the presence of the bicritical point and the first-order transition line. The magnetization plateau, on the other hand, closes after a KT transition only in the region $J_\times < J_\parallel$, in contrast to cases $J_\parallel=0.2$ and $0.55$. 
The suppression of the KT transition point on the side $J_\times>J_\parallel$ can be attributed to an increase in the downward slope of the critical field for the condensation of the singlet component, compared with the free hard-core bosons model, due to an enhancement in the relevance of the interaction terms, compared with the case $J_\parallel=0.55$. 
Although the transition lines in Fig. \ref{fig:otherjp} have been estimated after a finite-size scaling analysis, we can have a vivid representation of the phases and their transitions by
calculating the dimer correlations $\langle\mathbf{S}_{l,1}\cdot\mathbf{S}_{l,2}\rangle$ in a $J_\times$ scan, for fixed values of $h$, Fig. \ref{fig:otherjp}(c). In a $J_\times$ scan 
\cite{*[{}][{, and references therein.}] Jiang2023}, we consider a ladder in which the value of $J_\times$ increases linearly from the left to the right side, such that it covers a given range of $J_\times$, with a constant magnetic field $h$ along the chain. In Fig. \ref{fig:otherjp}(c), we also marked with an orange line the estimated values of the thermodynamic limit first-order transition points, as shown in the phase diagram for the respective values of $h$. We notice that $\langle\mathbf{S}_{l,1}\cdot\mathbf{S}_{l,2}\rangle=0.25$ on the right side of the transition, phase II; while $\langle\mathbf{S}_{l,1}\cdot\mathbf{S}_{l,2}\rangle$ ranges from $\approx -0.75$ for $h=0$ to $\approx 0.0$ for $h=2.43$. Thus, on the right side of the first-order transition line, we observe an effective spin-1 chain. In particular, for $m=0$, there is a gapped Haldane phase with a nontrivial topological ground state, in contrast to the trivial gapped $m=0$ state (PM) of the left side of the first-order transition line.

\subsection{Changing $J_\parallel$ with $J_\times$ fixed, and KT transition points}
Due to the ladder symmetry under the exchange of $J_\times$ and $J_\parallel$, and of the label of the spins in odd (or even) dimers,
the phase diagrams $h$ versus $J_\parallel$ for fixed $J_\times=0.55, 0.2,\text{ and }0.8$ exhibit the same transition lines shown in Figs. \ref{fig:fase} and \ref{fig:otherjp}. In particular, the expressions for the critical FP field shown in Fig. \ref{fig:freehfp} can be used to localize the bicritical point in the phase diagram $h$ versus $J_\parallel$ for fixed $J_\times$. The bicritical point appears whenever there is a crossing between solutions $h^s_c$ and $h^t_c$ as we vary $J_\times$ or $J_\parallel$,  with the second parameter constant. For fixed $J_\times$, from Fig.\ref{fig:freehfp}, we can assert the absence of the bicritical point for $J_\times<0.5$, and we anticipate the presence of a bicritical point at $J_\parallel=0.5$ and the field $h^s_c\vert_{J_{\parallel}=0.5}=h^t_c\vert_{J_{\parallel}=0.5}=1+2J_\times$, for $J_\times>0.5$. 

\begin{figure}
\includegraphics[width=0.35\textwidth]{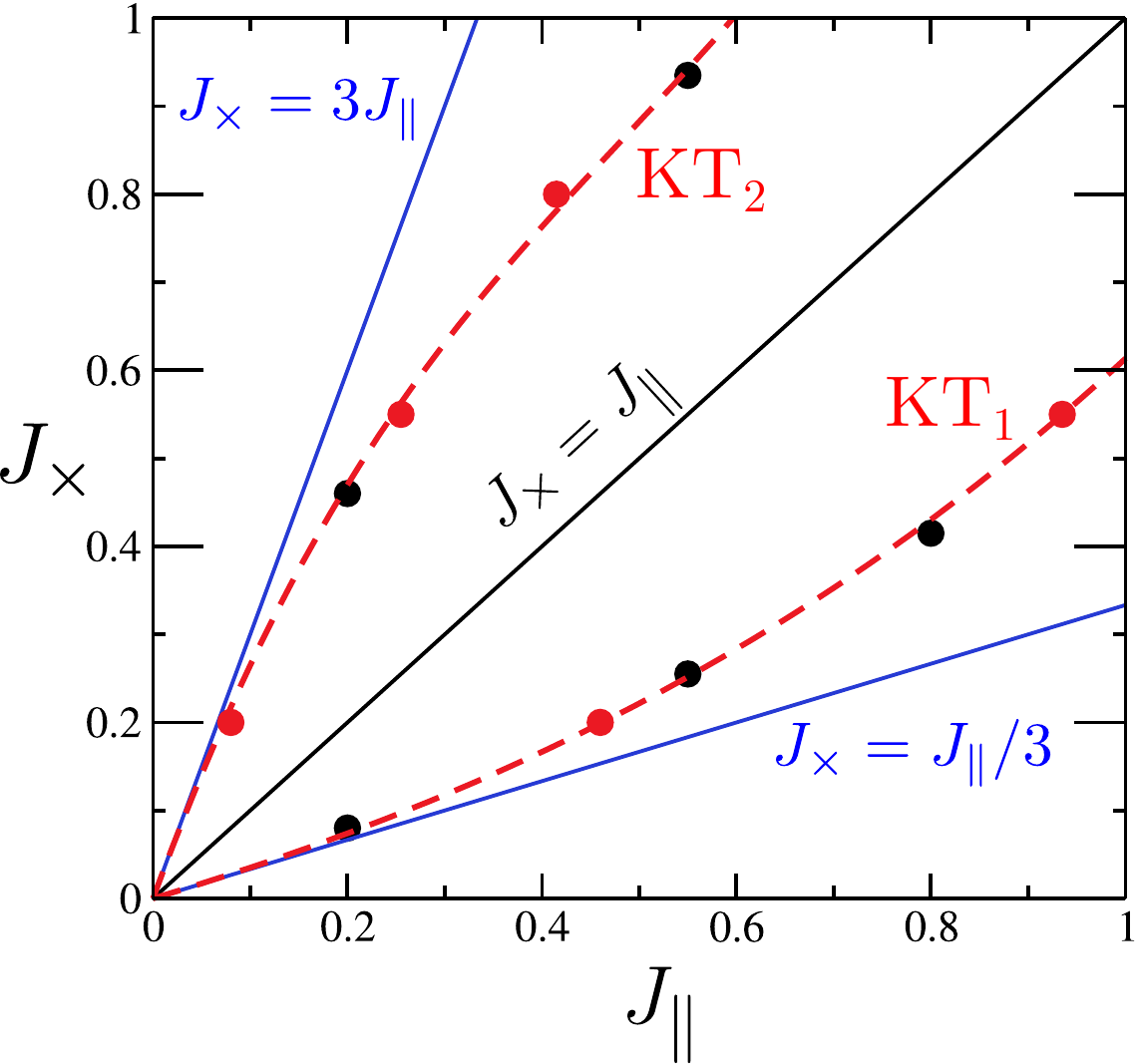}
\caption{The two lines of the KT transitions in the $J_\times$ versus $J_\parallel$ plane: the points marked by ($\bullet$) were calculated, while the points ({\color{red} $\bullet$}) were drawn by exploiting the symmetry of the ladder. Error bars are less than or equal to the size of the symbols. Full lines $J_\times=3J_\parallel$ and $J_\times=J_\parallel/3$ are the results of perturbation theory. Dashed lines are the best fittings to the numerical results of KT$_1$ and KT$_2$: $J_\parallel/3+0.16J_\parallel^2+0.12J_\parallel^3$ and 
$3J_\parallel-3.77J_\parallel^2+2.60J_\parallel^3$, respectively.}
\label{fig:jc}
\end{figure}

We can also exploit the symmetry of the chain to obtain a more general behavior for the KT transition points as a function of $J_\times$ and $J_\parallel$. In Fig. \ref{fig:jc}, the calculated points for $J_\parallel=0.2,~ 0.55,\text{ and }0.8$, which were shown in Figs. \ref{fig:fase} and \ref{fig:otherjp}, are supplemented by the corresponding symmetric points. We also show the prediction from perturbation theory\cite{Mila_1998}: $J_\times=3J_\parallel$ and $J_\times=J_\parallel/3$ for comparison. The curves for KT$_1$ and KT$_2$ can be well fitted considering corrections $J_\parallel^2$ and $J_\parallel^3$ to the analytical expression of perturbation: $J_\parallel/3+0.16J_\parallel^2+0.12J_\parallel^3$ for KT$_1$, and $3J_\parallel-3.77J_\parallel^2+2.60J_\parallel^3$ for KT$_2$.  

\section{Summary}
\label{sec:sum}

We investigated the frustrated spin-1/2 ladder in a magnetic field $h$ using DMRG and hard-core boson mapping. We have focused, in particular, on the quantum bicritical point, the first-order transition line, and the precise numerical determination of the Kosterlitz-Thouless 
transition points. We present numerical data from phase diagrams $h$ versus $J_\times$ with $J_\parallel=0.2,~0.55,\text{ and }0.8$, but we use symmetry arguments and boson mapping to discuss phase diagrams $h$ versus $J_\parallel$ with $J_\times$ fixed. 

The quantum bicritical point, which is the intersection of two second-order transition lines from the fully polarized phase to two distinct gapless phases, is observed at $(m=1,J_\times=0.5)$ in the phase diagram $h$ versus $J_\times$ for constant $J_\parallel>0.5$, and at $(m=1,J_\parallel=0.5)$ for constant $J_\times>0.5$. The two competing gapless phases
are characterized by the probability density for the occurrence of a singlet, 
$\langle n_s\rangle$, or a triplet with the rung state $S^z=0$, $\langle n_{t_0}\rangle$.
With the help of these quantities, we have shown that
one phase is composed of local singlets and $\ket{\uparrow\uparrow}$
triplets, while the other has only local triplets $\ket{\uparrow\uparrow}$ 
and triplets $\ket{t_0}$. In the last phase, the frustrated ladder is effectively described by spin-1 chain in a magnetic field. In particular, the transverse spin correlation functions show that the total dimer spin has an antiferromagnetic quasi-long-range order in the direction perpendicular to the magnetic field in the phase rich in triplets $\ket{t_0}$. The transition from one gapless phase to the other is of the first order kind in the field $h=h_f(J_\times)$, and the magnetization curve as a function of the magnetic field exhibits a jump in the transition between them. 

The first-order transition line $h_f(J_\times)$ is observed in the phase diagram for $J_\parallel=0.55\text{ and }0.8$, and is absent for $J_\parallel=0.2$. In fact, we argued that this line and the quantum bicritical point are not observed for $J_\parallel<0.5$, or for $J_\times<0.5$ in the phase diagram $h$ versus $J_\parallel$. 

The $h=h_f(J_\times)$ line starts at the quantum bicritical point, and to complete the whole physical picture, we have investigated the energy curves and phase coexistence along this line. Under a magnetic field $h=h_f$, the finite-size energy density curves exhibit two global minima corresponding to the magnetization states that bound the magnetization jump: $m_i^{(\text{jump})}$ and $m_f^{(\text{jump})}$, with unstable states between them. However, the scaling behavior of the energy density curves shows that the thermodynamic limit curve is flat for magnetizations $m_i^{(\text{jump})}<m<m_f^{(\text{jump})}$, and thus is a stable equilibrium curve. In fact, we attribute to the interface between the two coexisting phases the departure $~1/L$ of the finite-size energy curves from the stable thermodynamic limit curve. By calculating the distribution of singlets and triplets $\ket{t_0}$ along the ladder, we were able to reveal that the states in the range $m_i^{(\text{jump})}<m<m_f^{(\text{jump})}$ are
phase separated, and the two single-phase magnetizations [$m=m_i^{(\text{jump})}$ and $m=m_f^{(\text{jump})}$] coexist in distinct regions of the ladder. Following the magnitude of the magnetization jump, $\Delta m$, we have shown that $\Delta m\rightarrow 0$ as the bicritical point is approached along the line $h_f$. 

We have made a precise estimation of the Kosterlitz-Thouless (KT) transition points associated with the closing of the fractional $m=1/2$ plateau for $J_\parallel=0.2,~0.55\text{, and }0.80$ using the transverse spin correlation functions. 
For $J_\parallel=0.2\text{ and }0.55$, there is one KT point on each side of the phase diagram: $J_\times<J_\parallel$ and $J_\times>J_\parallel$; while for $J_\parallel=0.8$, there is only one in the region $J_\times<J_\parallel$. For $J_\parallel=0.8$, the first-order transition line $h_f$ crosses the plateau $m = 1 / 2$ so that there is a first-order transition from the plateau phase to the spin-1 phase. Finally, taking advantage of the ladder symmetry, we were able to estimate the curves of the KT points in the plane $J_\times$ versus $J_\parallel$.

We expect that our results will stimulate theoretical and experimental investigations in frustrated ladder systems, such as solid-state compounds or even correlated atoms in optical lattices, particularly focusing on the dynamical and thermal features associated with non-equilibrium states.

\section{ACKNOWLEDGMENTS}
We acknowledge the support from Coordenação de Aperfeiçoamento de Pessoal de Nível Superior (CAPES), Conselho Nacional de Desenvolvimento Científico e Tecnológico (CNPq), and Fundação de Amparo à Ciência e Tecnologia do Estado de Pernambuco (FACEPE), Brazilian agencies, including the PRONEX Program which is funded by CNPq and FACEPE, Grant No. APQ-0602-1.05/14. 

\begin{thebibliography}{52}%
\makeatletter
\providecommand \@ifxundefined [1]{%
 \@ifx{#1\undefined}
}%
\providecommand \@ifnum [1]{%
 \ifnum #1\expandafter \@firstoftwo
 \else \expandafter \@secondoftwo
 \fi
}%
\providecommand \@ifx [1]{%
 \ifx #1\expandafter \@firstoftwo
 \else \expandafter \@secondoftwo
 \fi
}%
\providecommand \natexlab [1]{#1}%
\providecommand \enquote  [1]{``#1''}%
\providecommand \bibnamefont  [1]{#1}%
\providecommand \bibfnamefont [1]{#1}%
\providecommand \citenamefont [1]{#1}%
\providecommand \href@noop [0]{\@secondoftwo}%
\providecommand \href [0]{\begingroup \@sanitize@url \@href}%
\providecommand \@href[1]{\@@startlink{#1}\@@href}%
\providecommand \@@href[1]{\endgroup#1\@@endlink}%
\providecommand \@sanitize@url [0]{\catcode `\\12\catcode `\$12\catcode
  `\&12\catcode `\#12\catcode `\^12\catcode `\_12\catcode `\%12\relax}%
\providecommand \@@startlink[1]{}%
\providecommand \@@endlink[0]{}%
\providecommand \url  [0]{\begingroup\@sanitize@url \@url }%
\providecommand \@url [1]{\endgroup\@href {#1}{\urlprefix }}%
\providecommand \urlprefix  [0]{URL }%
\providecommand \Eprint [0]{\href }%
\providecommand \doibase [0]{https://doi.org/}%
\providecommand \selectlanguage [0]{\@gobble}%
\providecommand \bibinfo  [0]{\@secondoftwo}%
\providecommand \bibfield  [0]{\@secondoftwo}%
\providecommand \translation [1]{[#1]}%
\providecommand \BibitemOpen [0]{}%
\providecommand \bibitemStop [0]{}%
\providecommand \bibitemNoStop [0]{.\EOS\space}%
\providecommand \EOS [0]{\spacefactor3000\relax}%
\providecommand \BibitemShut  [1]{\csname bibitem#1\endcsname}%
\let\auto@bib@innerbib\@empty
%</preamble>
\bibitem [{\citenamefont {Chaikin}\ and\ \citenamefont
  {Lubensky}(2000)}]{chaikin2000principles}%
  \BibitemOpen
  \bibfield  {author} {\bibinfo {author} {\bibfnamefont {P.}~\bibnamefont
  {Chaikin}}\ and\ \bibinfo {author} {\bibfnamefont {T.}~\bibnamefont
  {Lubensky}},\ }\href {https://books.google.com.br/books?id=P9YjNjzr9OIC}
  {\emph {\bibinfo {title} {Principles of Condensed Matter Physics}}}\
  (\bibinfo  {publisher} {Cambridge University Press},\ \bibinfo {year}
  {2000})\BibitemShut {NoStop}%
\bibitem [{\citenamefont {Sachdev}(2011)}]{sachdev2011quantum}%
  \BibitemOpen
  \bibfield  {author} {\bibinfo {author} {\bibfnamefont {S.}~\bibnamefont
  {Sachdev}},\ }\href {https://books.google.com.br/books?id=F3IkpxwpqSgC}
  {\emph {\bibinfo {title} {Quantum Phase Transitions}}}\ (\bibinfo
  {publisher} {Cambridge University Press},\ \bibinfo {year}
  {2011})\BibitemShut {NoStop}%
\bibitem [{\citenamefont {Continentino}(2017)}]{continentino_2017}%
  \BibitemOpen
  \bibfield  {author} {\bibinfo {author} {\bibfnamefont {M.}~\bibnamefont
  {Continentino}},\ }\href {https://doi.org/10.1017/CBO9781316576854} {\emph
  {\bibinfo {title} {Quantum Scaling in Many-Body Systems: An Approach to
  Quantum Phase Transitions}}},\ \bibinfo {edition} {2nd}\ ed.\ (\bibinfo
  {publisher} {Cambridge University Press},\ \bibinfo {year}
  {2017})\BibitemShut {NoStop}%
\bibitem [{\citenamefont {Lacroix}\ \emph {et~al.}(2011)\citenamefont
  {Lacroix}, \citenamefont {Mendels},\ and\ \citenamefont
  {Mila}}]{lacroix2011introduction}%
  \BibitemOpen
  \bibfield  {author} {\bibinfo {author} {\bibfnamefont {C.}~\bibnamefont
  {Lacroix}}, \bibinfo {author} {\bibfnamefont {P.}~\bibnamefont {Mendels}},\
  and\ \bibinfo {author} {\bibfnamefont {F.}~\bibnamefont {Mila}},\ }\href
  {https://books.google.com.br/books?id=utSV09ZuhOkC} {\emph {\bibinfo {title}
  {{Introduction to Frustrated Magnetism: Materials, Experiments, Theory}}}},\
  Springer Series in Solid-State Sciences\ (\bibinfo  {publisher} {Springer
  Berlin Heidelberg},\ \bibinfo {year} {2011})\BibitemShut {NoStop}%
\bibitem [{\citenamefont {Vojta}(2018)}]{Vojta2018}%
  \BibitemOpen
  \bibfield  {author} {\bibinfo {author} {\bibfnamefont {M.}~\bibnamefont
  {Vojta}},\ }\bibfield  {title} {\bibinfo {title} {{Frustration and quantum
  criticality}},\ }\href {https://doi.org/10.1088/1361-6633/aab6be} {\bibfield
  {journal} {\bibinfo  {journal} {Reports on Progress in Physics}\ }\textbf
  {\bibinfo {volume} {81}},\ \bibinfo {pages} {064501} (\bibinfo {year}
  {2018})}\BibitemShut {NoStop}%
\bibitem [{\citenamefont {Jim{\'{e}}nez}\ \emph {et~al.}(2021)\citenamefont
  {Jim{\'{e}}nez}, \citenamefont {Crone}, \citenamefont {Fogh}, \citenamefont
  {Zayed}, \citenamefont {Lortz}, \citenamefont {Pomjakushina}, \citenamefont
  {Conder}, \citenamefont {L{\"{a}}uchli}, \citenamefont {Weber}, \citenamefont
  {Wessel}, \citenamefont {Honecker}, \citenamefont {Normand}, \citenamefont
  {R{\"{u}}egg}, \citenamefont {Corboz}, \citenamefont {R{\o}nnow},\ and\
  \citenamefont {Mila}}]{Jimenez2021}%
  \BibitemOpen
  \bibfield  {author} {\bibinfo {author} {\bibfnamefont {J.~L.}\ \bibnamefont
  {Jim{\'{e}}nez}}, \bibinfo {author} {\bibfnamefont {S.~P.~G.}\ \bibnamefont
  {Crone}}, \bibinfo {author} {\bibfnamefont {E.}~\bibnamefont {Fogh}},
  \bibinfo {author} {\bibfnamefont {M.~E.}\ \bibnamefont {Zayed}}, \bibinfo
  {author} {\bibfnamefont {R.}~\bibnamefont {Lortz}}, \bibinfo {author}
  {\bibfnamefont {E.}~\bibnamefont {Pomjakushina}}, \bibinfo {author}
  {\bibfnamefont {K.}~\bibnamefont {Conder}}, \bibinfo {author} {\bibfnamefont
  {A.~M.}\ \bibnamefont {L{\"{a}}uchli}}, \bibinfo {author} {\bibfnamefont
  {L.}~\bibnamefont {Weber}}, \bibinfo {author} {\bibfnamefont
  {S.}~\bibnamefont {Wessel}}, \bibinfo {author} {\bibfnamefont
  {A.}~\bibnamefont {Honecker}}, \bibinfo {author} {\bibfnamefont
  {B.}~\bibnamefont {Normand}}, \bibinfo {author} {\bibfnamefont
  {C.}~\bibnamefont {R{\"{u}}egg}}, \bibinfo {author} {\bibfnamefont
  {P.}~\bibnamefont {Corboz}}, \bibinfo {author} {\bibfnamefont {H.~M.}\
  \bibnamefont {R{\o}nnow}},\ and\ \bibinfo {author} {\bibfnamefont
  {F.}~\bibnamefont {Mila}},\ }\bibfield  {title} {\bibinfo {title} {{A quantum
  magnetic analogue to the critical point of water}},\ }\href
  {https://doi.org/10.1038/s41586-021-03411-8} {\bibfield  {journal} {\bibinfo
  {journal} {Nature}\ }\textbf {\bibinfo {volume} {592}},\ \bibinfo {pages}
  {370} (\bibinfo {year} {2021})}\BibitemShut {NoStop}%
\bibitem [{\citenamefont {Stapmanns}\ \emph {et~al.}(2018)\citenamefont
  {Stapmanns}, \citenamefont {Corboz}, \citenamefont {Mila}, \citenamefont
  {Honecker}, \citenamefont {Normand},\ and\ \citenamefont
  {Wessel}}]{Stapmanns2018}%
  \BibitemOpen
  \bibfield  {author} {\bibinfo {author} {\bibfnamefont {J.}~\bibnamefont
  {Stapmanns}}, \bibinfo {author} {\bibfnamefont {P.}~\bibnamefont {Corboz}},
  \bibinfo {author} {\bibfnamefont {F.}~\bibnamefont {Mila}}, \bibinfo {author}
  {\bibfnamefont {A.}~\bibnamefont {Honecker}}, \bibinfo {author}
  {\bibfnamefont {B.}~\bibnamefont {Normand}},\ and\ \bibinfo {author}
  {\bibfnamefont {S.}~\bibnamefont {Wessel}},\ }\bibfield  {title} {\bibinfo
  {title} {{Thermal Critical Points and Quantum Critical End Point in the
  Frustrated Bilayer Heisenberg Antiferromagnet}},\ }\href
  {https://doi.org/10.1103/PhysRevLett.121.127201} {\bibfield  {journal}
  {\bibinfo  {journal} {Physical Review Letters}\ }\textbf {\bibinfo {volume}
  {121}},\ \bibinfo {pages} {127201} (\bibinfo {year} {2018})}\BibitemShut
  {NoStop}%
\bibitem [{\citenamefont {Tokiwa}\ \emph {et~al.}(2013)\citenamefont {Tokiwa},
  \citenamefont {Garst}, \citenamefont {Gegenwart}, \citenamefont {Bud'ko},\
  and\ \citenamefont {Canfield}}]{Tokiwa2013}%
  \BibitemOpen
  \bibfield  {author} {\bibinfo {author} {\bibfnamefont {Y.}~\bibnamefont
  {Tokiwa}}, \bibinfo {author} {\bibfnamefont {M.}~\bibnamefont {Garst}},
  \bibinfo {author} {\bibfnamefont {P.}~\bibnamefont {Gegenwart}}, \bibinfo
  {author} {\bibfnamefont {S.~L.}\ \bibnamefont {Bud'ko}},\ and\ \bibinfo
  {author} {\bibfnamefont {P.~C.}\ \bibnamefont {Canfield}},\ }\bibfield
  {title} {\bibinfo {title} {{Quantum Bicriticality in the Heavy-Fermion
  Metamagnet YbAgGe}},\ }\href {https://doi.org/10.1103/PhysRevLett.111.116401}
  {\bibfield  {journal} {\bibinfo  {journal} {Physical Review Letters}\
  }\textbf {\bibinfo {volume} {111}},\ \bibinfo {pages} {116401} (\bibinfo
  {year} {2013})}\BibitemShut {NoStop}%
\bibitem [{\citenamefont {Fisher}\ and\ \citenamefont
  {Nelson}(1974)}]{Fisher1974}%
  \BibitemOpen
  \bibfield  {author} {\bibinfo {author} {\bibfnamefont {M.~E.}\ \bibnamefont
  {Fisher}}\ and\ \bibinfo {author} {\bibfnamefont {D.~R.}\ \bibnamefont
  {Nelson}},\ }\bibfield  {title} {\bibinfo {title} {Spin flop, supersolids,
  and bicritical and tetracritical points},\ }\href
  {https://doi.org/10.1103/PhysRevLett.32.1350} {\bibfield  {journal} {\bibinfo
   {journal} {Physical Review Letters}\ }\textbf {\bibinfo {volume} {32}},\
  \bibinfo {pages} {1350} (\bibinfo {year} {1974})}\BibitemShut {NoStop}%
\bibitem [{\citenamefont {Kosterlitz}\ \emph {et~al.}(1976)\citenamefont
  {Kosterlitz}, \citenamefont {Nelson},\ and\ \citenamefont
  {Fisher}}]{Kosterlitz1976}%
  \BibitemOpen
  \bibfield  {author} {\bibinfo {author} {\bibfnamefont {J.~M.}\ \bibnamefont
  {Kosterlitz}}, \bibinfo {author} {\bibfnamefont {D.~R.}\ \bibnamefont
  {Nelson}},\ and\ \bibinfo {author} {\bibfnamefont {M.~E.}\ \bibnamefont
  {Fisher}},\ }\bibfield  {title} {\bibinfo {title} {Bicritical and
  tetracritical points in anisotropic antiferromagnetic systems},\ }\href
  {https://doi.org/10.1103/PhysRevB.13.412} {\bibfield  {journal} {\bibinfo
  {journal} {Physical Review B}\ }\textbf {\bibinfo {volume} {13}},\ \bibinfo
  {pages} {412} (\bibinfo {year} {1976})}\BibitemShut {NoStop}%
\bibitem [{\citenamefont {Morice}\ \emph {et~al.}(2017)\citenamefont {Morice},
  \citenamefont {Chandra}, \citenamefont {Rowley}, \citenamefont {Lonzarich},\
  and\ \citenamefont {Saxena}}]{Morice2017}%
  \BibitemOpen
  \bibfield  {author} {\bibinfo {author} {\bibfnamefont {C.}~\bibnamefont
  {Morice}}, \bibinfo {author} {\bibfnamefont {P.}~\bibnamefont {Chandra}},
  \bibinfo {author} {\bibfnamefont {S.~E.}\ \bibnamefont {Rowley}}, \bibinfo
  {author} {\bibfnamefont {G.}~\bibnamefont {Lonzarich}},\ and\ \bibinfo
  {author} {\bibfnamefont {S.~S.}\ \bibnamefont {Saxena}},\ }\bibfield  {title}
  {\bibinfo {title} {{Hidden fluctuations close to a quantum bicritical
  point}},\ }\bibfield  {journal} {\bibinfo  {journal} {Physical Review B}\
  }\textbf {\bibinfo {volume} {96}},\ \href
  {https://doi.org/10.1103/PhysRevB.96.245104} {10.1103/PhysRevB.96.245104}
  (\bibinfo {year} {2017})\BibitemShut {NoStop}%
\bibitem [{\citenamefont {Lopes}\ \emph
  {et~al.}(2020{\natexlab{a}})\citenamefont {Lopes}, \citenamefont {Barci},\
  and\ \citenamefont {Continentino}}]{Lopes2020}%
  \BibitemOpen
  \bibfield  {author} {\bibinfo {author} {\bibfnamefont {N.}~\bibnamefont
  {Lopes}}, \bibinfo {author} {\bibfnamefont {D.~G.}\ \bibnamefont {Barci}},\
  and\ \bibinfo {author} {\bibfnamefont {M.~A.}\ \bibnamefont {Continentino}},\
  }\bibfield  {title} {\bibinfo {title} {{Finite temperature effects in quantum
  systems with competing scalar orders}},\ }\href
  {https://doi.org/10.1088/1361-648X/ab9a7c} {\bibfield  {journal} {\bibinfo
  {journal} {Journal of Physics: Condensed Matter}\ }\textbf {\bibinfo {volume}
  {32}},\ \bibinfo {pages} {415601} (\bibinfo {year}
  {2020}{\natexlab{a}})}\BibitemShut {NoStop}%
\bibitem [{\citenamefont {Lopes}\ \emph
  {et~al.}(2020{\natexlab{b}})\citenamefont {Lopes}, \citenamefont
  {Continentino},\ and\ \citenamefont {Barci}}]{Lopes2020a}%
  \BibitemOpen
  \bibfield  {author} {\bibinfo {author} {\bibfnamefont {N.}~\bibnamefont
  {Lopes}}, \bibinfo {author} {\bibfnamefont {M.~A.}\ \bibnamefont
  {Continentino}},\ and\ \bibinfo {author} {\bibfnamefont {D.~G.}\ \bibnamefont
  {Barci}},\ }\bibfield  {title} {\bibinfo {title} {{One-loop effective
  potential for two-dimensional competing scalar order parameters}},\ }\href
  {https://doi.org/10.1016/j.physleta.2019.126095} {\bibfield  {journal}
  {\bibinfo  {journal} {Physics Letters A}\ }\textbf {\bibinfo {volume}
  {384}},\ \bibinfo {pages} {126095} (\bibinfo {year}
  {2020}{\natexlab{b}})}\BibitemShut {NoStop}%
\bibitem [{\citenamefont {Dagotto}(1999)}]{Dagotto1999}%
  \BibitemOpen
  \bibfield  {author} {\bibinfo {author} {\bibfnamefont {E.}~\bibnamefont
  {Dagotto}},\ }\bibfield  {title} {\bibinfo {title} {{Experiments on Ladders
  Reveal a Complex Interplay between a Spin-Gapped Normal State and
  Superconductivity}},\ }\href {https://doi.org/10.1088/0034-4885/62/11/202}
  {\bibfield  {journal} {\bibinfo  {journal} {Rep. Prog. Phys.}\ }\textbf
  {\bibinfo {volume} {62}},\ \bibinfo {pages} {22} (\bibinfo {year}
  {1999})}\BibitemShut {NoStop}%
\bibitem [{\citenamefont {White}\ \emph {et~al.}(1994)\citenamefont {White},
  \citenamefont {Noack},\ and\ \citenamefont {Scalapino}}]{White1994}%
  \BibitemOpen
  \bibfield  {author} {\bibinfo {author} {\bibfnamefont {S.~R.}\ \bibnamefont
  {White}}, \bibinfo {author} {\bibfnamefont {R.~M.}\ \bibnamefont {Noack}},\
  and\ \bibinfo {author} {\bibfnamefont {D.~J.}\ \bibnamefont {Scalapino}},\
  }\bibfield  {title} {\bibinfo {title} {Resonating valence bond theory of
  coupled heisenberg chains},\ }\href
  {https://doi.org/10.1103/PhysRevLett.73.886} {\bibfield  {journal} {\bibinfo
  {journal} {Physical Review Letters}\ }\textbf {\bibinfo {volume} {73}},\
  \bibinfo {pages} {886} (\bibinfo {year} {1994})}\BibitemShut {NoStop}%
\bibitem [{\citenamefont {Giamarchi}(2004)}]{giamarchi2003quantum}%
  \BibitemOpen
  \bibfield  {author} {\bibinfo {author} {\bibfnamefont {T.}~\bibnamefont
  {Giamarchi}},\ }\href@noop {} {\emph {\bibinfo {title} {Quantum physics in
  one dimension}}}\ (\bibinfo  {publisher} {Clarendon Press, Oxford},\ \bibinfo
  {year} {2004})\BibitemShut {NoStop}%
\bibitem [{\citenamefont {Chitra}\ and\ \citenamefont
  {Giamarchi}(1997)}]{PhysRevB.55.58}%
  \BibitemOpen
  \bibfield  {author} {\bibinfo {author} {\bibfnamefont {R.}~\bibnamefont
  {Chitra}}\ and\ \bibinfo {author} {\bibfnamefont {T.}~\bibnamefont
  {Giamarchi}},\ }\bibfield  {title} {\bibinfo {title} {{Critical properties of
  gapped spin-chains and ladders in a magnetic field}},\ }\href
  {https://doi.org/10.1103/PhysRevB.55.5816} {\bibfield  {journal} {\bibinfo
  {journal} {Physical Review B}\ }\textbf {\bibinfo {volume} {55}},\ \bibinfo
  {pages} {5816} (\bibinfo {year} {1997})}\BibitemShut {NoStop}%
\bibitem [{\citenamefont {Giamarchi}\ and\ \citenamefont
  {Tsvelik}(1999)}]{PhysRevB.59.11398}%
  \BibitemOpen
  \bibfield  {author} {\bibinfo {author} {\bibfnamefont {T.}~\bibnamefont
  {Giamarchi}}\ and\ \bibinfo {author} {\bibfnamefont {A.~M.}\ \bibnamefont
  {Tsvelik}},\ }\bibfield  {title} {\bibinfo {title} {{Coupled ladders in a
  magnetic field}},\ }\href {https://doi.org/10.1103/PhysRevB.59.11398}
  {\bibfield  {journal} {\bibinfo  {journal} {Physical Review B}\ }\textbf
  {\bibinfo {volume} {59}},\ \bibinfo {pages} {11398} (\bibinfo {year}
  {1999})}\BibitemShut {NoStop}%
\bibitem [{\citenamefont {Hikihara}\ and\ \citenamefont
  {Furusaki}(2001)}]{Hikihara2001}%
  \BibitemOpen
  \bibfield  {author} {\bibinfo {author} {\bibfnamefont {T.}~\bibnamefont
  {Hikihara}}\ and\ \bibinfo {author} {\bibfnamefont {A.}~\bibnamefont
  {Furusaki}},\ }\bibfield  {title} {\bibinfo {title} {{Spin correlations in
  the two-leg antiferromagnetic ladder in a magnetic field}},\ }\href
  {https://doi.org/10.1103/PhysRevB.63.134438} {\bibfield  {journal} {\bibinfo
  {journal} {Physical Review B}\ }\textbf {\bibinfo {volume} {63}},\ \bibinfo
  {pages} {134438} (\bibinfo {year} {2001})}\BibitemShut {NoStop}%
\bibitem [{\citenamefont {R{\"{u}}egg}\ \emph {et~al.}(2008)\citenamefont
  {R{\"{u}}egg}, \citenamefont {Kiefer}, \citenamefont {Thielemann},
  \citenamefont {McMorrow}, \citenamefont {Zapf}, \citenamefont {Normand},
  \citenamefont {Zvonarev}, \citenamefont {Bouillot}, \citenamefont {Kollath},
  \citenamefont {Giamarchi}, \citenamefont {Capponi}, \citenamefont
  {Poilblanc}, \citenamefont {Biner},\ and\ \citenamefont
  {Kr{\"{a}}mer}}]{Ruegg2008}%
  \BibitemOpen
  \bibfield  {author} {\bibinfo {author} {\bibfnamefont {C.}~\bibnamefont
  {R{\"{u}}egg}}, \bibinfo {author} {\bibfnamefont {K.}~\bibnamefont {Kiefer}},
  \bibinfo {author} {\bibfnamefont {B.}~\bibnamefont {Thielemann}}, \bibinfo
  {author} {\bibfnamefont {D.~F.}\ \bibnamefont {McMorrow}}, \bibinfo {author}
  {\bibfnamefont {V.}~\bibnamefont {Zapf}}, \bibinfo {author} {\bibfnamefont
  {B.}~\bibnamefont {Normand}}, \bibinfo {author} {\bibfnamefont {M.~B.}\
  \bibnamefont {Zvonarev}}, \bibinfo {author} {\bibfnamefont {P.}~\bibnamefont
  {Bouillot}}, \bibinfo {author} {\bibfnamefont {C.}~\bibnamefont {Kollath}},
  \bibinfo {author} {\bibfnamefont {T.}~\bibnamefont {Giamarchi}}, \bibinfo
  {author} {\bibfnamefont {S.}~\bibnamefont {Capponi}}, \bibinfo {author}
  {\bibfnamefont {D.}~\bibnamefont {Poilblanc}}, \bibinfo {author}
  {\bibfnamefont {D.}~\bibnamefont {Biner}},\ and\ \bibinfo {author}
  {\bibfnamefont {K.~W.}\ \bibnamefont {Kr{\"{a}}mer}},\ }\bibfield  {title}
  {\bibinfo {title} {{Thermodynamics of the Spin Luttinger Liquid in a Model
  Ladder Material}},\ }\href {https://doi.org/10.1103/PhysRevLett.101.247202}
  {\bibfield  {journal} {\bibinfo  {journal} {Physical Review Letters}\
  }\textbf {\bibinfo {volume} {101}},\ \bibinfo {pages} {247202} (\bibinfo
  {year} {2008})}\BibitemShut {NoStop}%
\bibitem [{\citenamefont {Klanj{\v{s}}ek}\ \emph {et~al.}(2008)\citenamefont
  {Klanj{\v{s}}ek}, \citenamefont {Mayaffre}, \citenamefont {Berthier},
  \citenamefont {Horvati{\'{c}}}, \citenamefont {Chiari}, \citenamefont
  {Piovesana}, \citenamefont {Bouillot}, \citenamefont {Kollath}, \citenamefont
  {Orignac}, \citenamefont {Citro},\ and\ \citenamefont
  {Giamarchi}}]{Klanjsek2008}%
  \BibitemOpen
  \bibfield  {author} {\bibinfo {author} {\bibfnamefont {M.}~\bibnamefont
  {Klanj{\v{s}}ek}}, \bibinfo {author} {\bibfnamefont {H.}~\bibnamefont
  {Mayaffre}}, \bibinfo {author} {\bibfnamefont {C.}~\bibnamefont {Berthier}},
  \bibinfo {author} {\bibfnamefont {M.}~\bibnamefont {Horvati{\'{c}}}},
  \bibinfo {author} {\bibfnamefont {B.}~\bibnamefont {Chiari}}, \bibinfo
  {author} {\bibfnamefont {O.}~\bibnamefont {Piovesana}}, \bibinfo {author}
  {\bibfnamefont {P.}~\bibnamefont {Bouillot}}, \bibinfo {author}
  {\bibfnamefont {C.}~\bibnamefont {Kollath}}, \bibinfo {author} {\bibfnamefont
  {E.}~\bibnamefont {Orignac}}, \bibinfo {author} {\bibfnamefont
  {R.}~\bibnamefont {Citro}},\ and\ \bibinfo {author} {\bibfnamefont
  {T.}~\bibnamefont {Giamarchi}},\ }\bibfield  {title} {\bibinfo {title}
  {{Controlling Luttinger Liquid Physics in Spin Ladders under a Magnetic
  Field}},\ }\href {https://doi.org/10.1103/PhysRevLett.101.137207} {\bibfield
  {journal} {\bibinfo  {journal} {Physical Review Letters}\ }\textbf {\bibinfo
  {volume} {101}},\ \bibinfo {pages} {137207} (\bibinfo {year}
  {2008})}\BibitemShut {NoStop}%
\bibitem [{\citenamefont {Thielemann}\ \emph {et~al.}(2009)\citenamefont
  {Thielemann}, \citenamefont {R{\"{u}}egg}, \citenamefont {R{\o}nnow},
  \citenamefont {L{\"{a}}uchli}, \citenamefont {Caux}, \citenamefont {Normand},
  \citenamefont {Biner}, \citenamefont {Kr{\"{a}}mer}, \citenamefont
  {G{\"{u}}del}, \citenamefont {Stahn}, \citenamefont {Habicht}, \citenamefont
  {Kiefer}, \citenamefont {Boehm}, \citenamefont {McMorrow},\ and\
  \citenamefont {Mesot}}]{Thielemann2009}%
  \BibitemOpen
  \bibfield  {author} {\bibinfo {author} {\bibfnamefont {B.}~\bibnamefont
  {Thielemann}}, \bibinfo {author} {\bibfnamefont {C.}~\bibnamefont
  {R{\"{u}}egg}}, \bibinfo {author} {\bibfnamefont {H.~M.}\ \bibnamefont
  {R{\o}nnow}}, \bibinfo {author} {\bibfnamefont {A.~M.}\ \bibnamefont
  {L{\"{a}}uchli}}, \bibinfo {author} {\bibfnamefont {J.~S.}\ \bibnamefont
  {Caux}}, \bibinfo {author} {\bibfnamefont {B.}~\bibnamefont {Normand}},
  \bibinfo {author} {\bibfnamefont {D.}~\bibnamefont {Biner}}, \bibinfo
  {author} {\bibfnamefont {K.~W.}\ \bibnamefont {Kr{\"{a}}mer}}, \bibinfo
  {author} {\bibfnamefont {H.~U.}\ \bibnamefont {G{\"{u}}del}}, \bibinfo
  {author} {\bibfnamefont {J.}~\bibnamefont {Stahn}}, \bibinfo {author}
  {\bibfnamefont {K.}~\bibnamefont {Habicht}}, \bibinfo {author} {\bibfnamefont
  {K.}~\bibnamefont {Kiefer}}, \bibinfo {author} {\bibfnamefont
  {M.}~\bibnamefont {Boehm}}, \bibinfo {author} {\bibfnamefont {D.~F.}\
  \bibnamefont {McMorrow}},\ and\ \bibinfo {author} {\bibfnamefont
  {J.}~\bibnamefont {Mesot}},\ }\bibfield  {title} {\bibinfo {title} {{Direct
  observation of magnon fractionalization in the quantum spin ladder}},\ }\href
  {https://doi.org/10.1103/PhysRevLett.102.107204} {\bibfield  {journal}
  {\bibinfo  {journal} {Physical Review Letters}\ }\textbf {\bibinfo {volume}
  {102}},\ \bibinfo {pages} {1} (\bibinfo {year} {2009})}\BibitemShut {NoStop}%
\bibitem [{\citenamefont {Gelfand}(1991)}]{Gelfand1991}%
  \BibitemOpen
  \bibfield  {author} {\bibinfo {author} {\bibfnamefont {M.~P.}\ \bibnamefont
  {Gelfand}},\ }\bibfield  {title} {\bibinfo {title} {Linked-tetrahedra spin
  chain: Exact ground state and excitations},\ }\href
  {https://doi.org/10.1103/PhysRevB.43.8644} {\bibfield  {journal} {\bibinfo
  {journal} {Physical Review B}\ }\textbf {\bibinfo {volume} {43}},\ \bibinfo
  {pages} {8644} (\bibinfo {year} {1991})}\BibitemShut {NoStop}%
\bibitem [{\citenamefont {White}(1996)}]{White1996}%
  \BibitemOpen
  \bibfield  {author} {\bibinfo {author} {\bibfnamefont {S.~R.}\ \bibnamefont
  {White}},\ }\bibfield  {title} {\bibinfo {title} {Equivalence of the
  antiferromagnetic heisenberg ladder to a single s =1 chain},\ }\href
  {https://doi.org/10.1103/PhysRevB.53.52} {\bibfield  {journal} {\bibinfo
  {journal} {Physical Review B}\ }\textbf {\bibinfo {volume} {53}},\ \bibinfo
  {pages} {52} (\bibinfo {year} {1996})}\BibitemShut {NoStop}%
\bibitem [{\citenamefont {Honecker}\ \emph {et~al.}(2000)\citenamefont
  {Honecker}, \citenamefont {Mila},\ and\ \citenamefont
  {Troyer}}]{Honecker_2000}%
  \BibitemOpen
  \bibfield  {author} {\bibinfo {author} {\bibfnamefont {A.}~\bibnamefont
  {Honecker}}, \bibinfo {author} {\bibfnamefont {F.}~\bibnamefont {Mila}},\
  and\ \bibinfo {author} {\bibfnamefont {M.}~\bibnamefont {Troyer}},\
  }\bibfield  {title} {\bibinfo {title} {Magnetization plateaux and jumps in a
  class of frustrated ladders: A simple route to a complex behaviour},\ }\href
  {https://doi.org/10.1007/s100510051120} {\bibfield  {journal} {\bibinfo
  {journal} {The European Physical Journal B}\ }\textbf {\bibinfo {volume}
  {15}},\ \bibinfo {pages} {227–233} (\bibinfo {year} {2000})}\BibitemShut
  {NoStop}%
\bibitem [{\citenamefont {Mila}(1998)}]{Mila_1998}%
  \BibitemOpen
  \bibfield  {author} {\bibinfo {author} {\bibfnamefont {F.}~\bibnamefont
  {Mila}},\ }\bibfield  {title} {\bibinfo {title} {Ladders in a magnetic field:
  a strong coupling approach},\ }\href {https://doi.org/10.1007/s100510050542}
  {\bibfield  {journal} {\bibinfo  {journal} {The European Physical Journal B}\
  }\textbf {\bibinfo {volume} {6}},\ \bibinfo {pages} {201–205} (\bibinfo
  {year} {1998})}\BibitemShut {NoStop}%
\bibitem [{\citenamefont {Tonegawa}\ \emph {et~al.}(1998)\citenamefont
  {Tonegawa}, \citenamefont {Nishida},\ and\ \citenamefont
  {Kaburagi}}]{Tonegawa1998}%
  \BibitemOpen
  \bibfield  {author} {\bibinfo {author} {\bibfnamefont {T.}~\bibnamefont
  {Tonegawa}}, \bibinfo {author} {\bibfnamefont {T.}~\bibnamefont {Nishida}},\
  and\ \bibinfo {author} {\bibfnamefont {M.}~\bibnamefont {Kaburagi}},\
  }\bibfield  {title} {\bibinfo {title} {{Ground-state magnetization curve of a
  generalized spin-1/2 ladder}},\ }\href
  {https://doi.org/10.1016/S0921-4526(97)00937-X} {\bibfield  {journal}
  {\bibinfo  {journal} {Physica B: Condensed Matter}\ }\textbf {\bibinfo
  {volume} {246-247}},\ \bibinfo {pages} {368} (\bibinfo {year}
  {1998})}\BibitemShut {NoStop}%
\bibitem [{\citenamefont {Totsuka}(1998)}]{Totsuka1998}%
  \BibitemOpen
  \bibfield  {author} {\bibinfo {author} {\bibfnamefont {K.}~\bibnamefont
  {Totsuka}},\ }\bibfield  {title} {\bibinfo {title} {{Magnetization plateau in
  the S=1/2 Heinsenberg spin chain with next-nearest-neighbor and alternating
  neare}},\ }\href {https://doi.org/10.1103/PhysRevB.57.3454} {\bibfield
  {journal} {\bibinfo  {journal} {Physical Review B}\ }\textbf {\bibinfo
  {volume} {57}},\ \bibinfo {pages} {3454} (\bibinfo {year}
  {1998})}\BibitemShut {NoStop}%
\bibitem [{\citenamefont {Fouet}\ \emph {et~al.}(2006)\citenamefont {Fouet},
  \citenamefont {Mila}, \citenamefont {Clarke}, \citenamefont {Youk},
  \citenamefont {Tchernyshyov}, \citenamefont {Fendley},\ and\ \citenamefont
  {Noack}}]{Fouet2006}%
  \BibitemOpen
  \bibfield  {author} {\bibinfo {author} {\bibfnamefont {J.-B.}\ \bibnamefont
  {Fouet}}, \bibinfo {author} {\bibfnamefont {F.}~\bibnamefont {Mila}},
  \bibinfo {author} {\bibfnamefont {D.}~\bibnamefont {Clarke}}, \bibinfo
  {author} {\bibfnamefont {H.}~\bibnamefont {Youk}}, \bibinfo {author}
  {\bibfnamefont {O.}~\bibnamefont {Tchernyshyov}}, \bibinfo {author}
  {\bibfnamefont {P.}~\bibnamefont {Fendley}},\ and\ \bibinfo {author}
  {\bibfnamefont {R.~M.}\ \bibnamefont {Noack}},\ }\bibfield  {title} {\bibinfo
  {title} {{Condensation of magnons and spinons in a frustrated ladder}},\
  }\href {https://doi.org/10.1103/PhysRevB.73.214405} {\bibfield  {journal}
  {\bibinfo  {journal} {Physical Review B}\ }\textbf {\bibinfo {volume} {73}},\
  \bibinfo {pages} {214405} (\bibinfo {year} {2006})}\BibitemShut {NoStop}%
\bibitem [{\citenamefont {Penc}\ \emph {et~al.}(2007)\citenamefont {Penc},
  \citenamefont {Fouet}, \citenamefont {Miyahara}, \citenamefont
  {Tchernyshyov},\ and\ \citenamefont {Mila}}]{Penc2007}%
  \BibitemOpen
  \bibfield  {author} {\bibinfo {author} {\bibfnamefont {K.}~\bibnamefont
  {Penc}}, \bibinfo {author} {\bibfnamefont {J.-B.}\ \bibnamefont {Fouet}},
  \bibinfo {author} {\bibfnamefont {S.}~\bibnamefont {Miyahara}}, \bibinfo
  {author} {\bibfnamefont {O.}~\bibnamefont {Tchernyshyov}},\ and\ \bibinfo
  {author} {\bibfnamefont {F.}~\bibnamefont {Mila}},\ }\bibfield  {title}
  {\bibinfo {title} {{Ising Phases of Heisenberg Ladders in a Magnetic
  Field}},\ }\href {https://doi.org/10.1103/PhysRevLett.99.117201} {\bibfield
  {journal} {\bibinfo  {journal} {Physical Review Letters}\ }\textbf {\bibinfo
  {volume} {99}},\ \bibinfo {pages} {117201} (\bibinfo {year}
  {2007})}\BibitemShut {NoStop}%
\bibitem [{\citenamefont {Michaud}\ \emph {et~al.}(2010)\citenamefont
  {Michaud}, \citenamefont {Coletta}, \citenamefont {Manmana}, \citenamefont
  {Picon},\ and\ \citenamefont {Mila}}]{Michaud2010}%
  \BibitemOpen
  \bibfield  {author} {\bibinfo {author} {\bibfnamefont {F.}~\bibnamefont
  {Michaud}}, \bibinfo {author} {\bibfnamefont {T.}~\bibnamefont {Coletta}},
  \bibinfo {author} {\bibfnamefont {S.~R.}\ \bibnamefont {Manmana}}, \bibinfo
  {author} {\bibfnamefont {J.-d.}\ \bibnamefont {Picon}},\ and\ \bibinfo
  {author} {\bibfnamefont {F.}~\bibnamefont {Mila}},\ }\bibfield  {title}
  {\bibinfo {title} {{Frustration-induced plateaus in S = 1/2 Heisenberg spin
  ladders}},\ }\href {https://doi.org/10.1103/PhysRevB.81.014407} {\bibfield
  {journal} {\bibinfo  {journal} {Physical Review B}\ }\textbf {\bibinfo
  {volume} {81}},\ \bibinfo {pages} {014407} (\bibinfo {year}
  {2010})}\BibitemShut {NoStop}%
\bibitem [{\citenamefont {Kosterlitz}(2017)}]{nobelkosterlitz}%
  \BibitemOpen
  \bibfield  {author} {\bibinfo {author} {\bibfnamefont {J.~M.}\ \bibnamefont
  {Kosterlitz}},\ }\bibfield  {title} {\bibinfo {title} {{Nobel Lecture:
  Topological defects and phase transitions}},\ }\href
  {https://doi.org/10.1103/RevModPhys.89.040501} {\bibfield  {journal}
  {\bibinfo  {journal} {Reviews of Modern Physics}\ }\textbf {\bibinfo {volume}
  {89}},\ \bibinfo {pages} {040501} (\bibinfo {year} {2017})}\BibitemShut
  {NoStop}%
\bibitem [{\citenamefont {Kosterlitz}\ and\ \citenamefont
  {Thouless}(1973)}]{Kosterlitz1973}%
  \BibitemOpen
  \bibfield  {author} {\bibinfo {author} {\bibfnamefont {J.~M.}\ \bibnamefont
  {Kosterlitz}}\ and\ \bibinfo {author} {\bibfnamefont {D.~J.}\ \bibnamefont
  {Thouless}},\ }\bibfield  {title} {\bibinfo {title} {{Ordering, metastability
  and phase transitions in two-dimensional systems}},\ }\href
  {https://doi.org/10.1088/0022-3719/6/7/010} {\bibfield  {journal} {\bibinfo
  {journal} {Journal of Physics C: Solid State Physics}\ }\textbf {\bibinfo
  {volume} {6}},\ \bibinfo {pages} {1181} (\bibinfo {year} {1973})}\BibitemShut
  {NoStop}%
\bibitem [{\citenamefont {Xavier}\ \emph {et~al.}(2022)\citenamefont {Xavier},
  \citenamefont {Pereira}, \citenamefont {Nunes},\ and\ \citenamefont
  {Plascak}}]{Xavier2022}%
  \BibitemOpen
  \bibfield  {author} {\bibinfo {author} {\bibfnamefont {J.~C.}\ \bibnamefont
  {Xavier}}, \bibinfo {author} {\bibfnamefont {R.~G.}\ \bibnamefont {Pereira}},
  \bibinfo {author} {\bibfnamefont {M.~E.~S.}\ \bibnamefont {Nunes}},\ and\
  \bibinfo {author} {\bibfnamefont {J.~A.}\ \bibnamefont {Plascak}},\
  }\bibfield  {title} {\bibinfo {title} {{Coexistence of spontaneous
  dimerization and magnetic order in a transverse-field Ising ladder with
  four-spin interactions}},\ }\href
  {https://doi.org/10.1103/PhysRevB.105.024430} {\bibfield  {journal} {\bibinfo
   {journal} {Physical Review B}\ }\textbf {\bibinfo {volume} {105}},\ \bibinfo
  {pages} {024430} (\bibinfo {year} {2022})}\BibitemShut {NoStop}%
\bibitem [{\citenamefont {Nayak}\ \emph {et~al.}(2020)\citenamefont {Nayak},
  \citenamefont {Blosser}, \citenamefont {Zheludev},\ and\ \citenamefont
  {Mila}}]{Nayak2020}%
  \BibitemOpen
  \bibfield  {author} {\bibinfo {author} {\bibfnamefont {M.}~\bibnamefont
  {Nayak}}, \bibinfo {author} {\bibfnamefont {D.}~\bibnamefont {Blosser}},
  \bibinfo {author} {\bibfnamefont {A.}~\bibnamefont {Zheludev}},\ and\
  \bibinfo {author} {\bibfnamefont {F.}~\bibnamefont {Mila}},\ }\bibfield
  {title} {\bibinfo {title} {{Magnetic-Field-Induced Bound States in Spin-1/2
  Ladders}},\ }\href {https://doi.org/10.1103/PhysRevLett.124.087203}
  {\bibfield  {journal} {\bibinfo  {journal} {Physical Review Letters}\
  }\textbf {\bibinfo {volume} {124}},\ \bibinfo {pages} {087203} (\bibinfo
  {year} {2020})}\BibitemShut {NoStop}%
\bibitem [{\citenamefont {do~Nascimento-Junior}\ and\ \citenamefont
  {Montenegro-Filho}(2019)}]{PhysRevB.99.064404}%
  \BibitemOpen
  \bibfield  {author} {\bibinfo {author} {\bibfnamefont {A.~M.}\ \bibnamefont
  {do~Nascimento-Junior}}\ and\ \bibinfo {author} {\bibfnamefont {R.~R.}\
  \bibnamefont {Montenegro-Filho}},\ }\bibfield  {title} {\bibinfo {title}
  {Magnetic phase separation in a frustrated ferrimagnetic chain under a
  magnetic field},\ }\href {https://doi.org/10.1103/PhysRevB.99.064404}
  {\bibfield  {journal} {\bibinfo  {journal} {Phys. Rev. B}\ }\textbf {\bibinfo
  {volume} {99}},\ \bibinfo {pages} {064404} (\bibinfo {year}
  {2019})}\BibitemShut {NoStop}%
\bibitem [{\citenamefont {Montenegro-Filho}\ \emph {et~al.}(2020)\citenamefont
  {Montenegro-Filho}, \citenamefont {Matias},\ and\ \citenamefont
  {Coutinho-Filho}}]{MontenegroFilho2020}%
  \BibitemOpen
  \bibfield  {author} {\bibinfo {author} {\bibfnamefont {R.~R.}\ \bibnamefont
  {Montenegro-Filho}}, \bibinfo {author} {\bibfnamefont {F.~S.}\ \bibnamefont
  {Matias}},\ and\ \bibinfo {author} {\bibfnamefont {M.~D.}\ \bibnamefont
  {Coutinho-Filho}},\ }\bibfield  {title} {\bibinfo {title} {Topology of
  many-body edge and extended quantum states in an open spin chain: 1/3
  plateau, kosterlitz-thouless transition, and luttinger liquid},\ }\href
  {https://doi.org/10.1103/PhysRevB.102.035137} {\bibfield  {journal} {\bibinfo
   {journal} {Phys. Rev. B}\ }\textbf {\bibinfo {volume} {102}},\ \bibinfo
  {pages} {035137} (\bibinfo {year} {2020})}\BibitemShut {NoStop}%
\bibitem [{\citenamefont {Veríssimo}\ \emph {et~al.}(2019)\citenamefont
  {Veríssimo}, \citenamefont {Pereira}, \citenamefont {Strečka},\ and\
  \citenamefont {Lyra}}]{Verissimo2019}%
  \BibitemOpen
  \bibfield  {author} {\bibinfo {author} {\bibfnamefont {L.~M.}\ \bibnamefont
  {Veríssimo}}, \bibinfo {author} {\bibfnamefont {M.~S.~S.}\ \bibnamefont
  {Pereira}}, \bibinfo {author} {\bibfnamefont {J.}~\bibnamefont {Strečka}},\
  and\ \bibinfo {author} {\bibfnamefont {M.~L.}\ \bibnamefont {Lyra}},\
  }\bibfield  {title} {\bibinfo {title} {Kosterlitz-thouless and gaussian
  criticalities in a mixed spin-(1/2,5/2,1/2) heisenberg branched chain with
  exchange anisotropy},\ }\href {https://doi.org/10.1103/PhysRevB.99.134408}
  {\bibfield  {journal} {\bibinfo  {journal} {Physical Review B}\ }\textbf
  {\bibinfo {volume} {99}},\ \bibinfo {pages} {134408} (\bibinfo {year}
  {2019})}\BibitemShut {NoStop}%
\bibitem [{\citenamefont {Karl'ová}\ \emph {et~al.}(2019)\citenamefont
  {Karl'ová}, \citenamefont {Strečka},\ and\ \citenamefont
  {Lyra}}]{Karlova2019}%
  \BibitemOpen
  \bibfield  {author} {\bibinfo {author} {\bibfnamefont {K.}~\bibnamefont
  {Karl'ová}}, \bibinfo {author} {\bibfnamefont {J.}~\bibnamefont
  {Strečka}},\ and\ \bibinfo {author} {\bibfnamefont {M.~L.}\ \bibnamefont
  {Lyra}},\ }\bibfield  {title} {\bibinfo {title} {Breakdown of intermediate
  one-half magnetization plateau of spin-1/2 ising-heisenberg and heisenberg
  branched chains at triple and kosterlitz-thouless critical points},\ }\href
  {https://doi.org/10.1103/PhysRevE.100.042127} {\bibfield  {journal} {\bibinfo
   {journal} {Physical Review E}\ }\textbf {\bibinfo {volume} {100}},\ \bibinfo
  {pages} {042127} (\bibinfo {year} {2019})}\BibitemShut {NoStop}%
\bibitem [{\citenamefont {Sakai}\ and\ \citenamefont
  {Yamamoto}(1999)}]{YamamotoPRB99}%
  \BibitemOpen
  \bibfield  {author} {\bibinfo {author} {\bibfnamefont {T.}~\bibnamefont
  {Sakai}}\ and\ \bibinfo {author} {\bibfnamefont {S.}~\bibnamefont
  {Yamamoto}},\ }\bibfield  {title} {\bibinfo {title} {Critical behavior of
  anisotropic heisenberg mixed-spin chains in a field},\ }\href
  {https://doi.org/10.1103/PhysRevB.60.4053} {\bibfield  {journal} {\bibinfo
  {journal} {Physical Review B}\ }\textbf {\bibinfo {volume} {60}},\ \bibinfo
  {pages} {4053} (\bibinfo {year} {1999})}\BibitemShut {NoStop}%
\bibitem [{\citenamefont {Schollw{\"o}ck}(2005)}]{Schollw_ck_2005}%
  \BibitemOpen
  \bibfield  {author} {\bibinfo {author} {\bibfnamefont {U.}~\bibnamefont
  {Schollw{\"o}ck}},\ }\bibfield  {title} {\bibinfo {title} {The density-matrix
  renormalization group},\ }\href {https://doi.org/10.1103/RevModPhys.77.259}
  {\bibfield  {journal} {\bibinfo  {journal} {Rev. Mod. Phys.}\ }\textbf
  {\bibinfo {volume} {77}},\ \bibinfo {pages} {259} (\bibinfo {year}
  {2005})}\BibitemShut {NoStop}%
\bibitem [{\citenamefont {White}(1992)}]{PhysRevLett.69.2863}%
  \BibitemOpen
  \bibfield  {author} {\bibinfo {author} {\bibfnamefont {S.~R.}\ \bibnamefont
  {White}},\ }\bibfield  {title} {\bibinfo {title} {Density matrix formulation
  for quantum renormalization groups},\ }\href
  {https://doi.org/10.1103/PhysRevLett.69.2863} {\bibfield  {journal} {\bibinfo
   {journal} {Phys. Rev. Lett.}\ }\textbf {\bibinfo {volume} {69}},\ \bibinfo
  {pages} {2863} (\bibinfo {year} {1992})}\BibitemShut {NoStop}%
\bibitem [{\citenamefont {White}(1993)}]{PhysRevB.48.10345}%
  \BibitemOpen
  \bibfield  {author} {\bibinfo {author} {\bibfnamefont {S.~R.}\ \bibnamefont
  {White}},\ }\bibfield  {title} {\bibinfo {title} {Density-matrix algorithms
  for quantum renormalization groups},\ }\href
  {https://doi.org/10.1103/PhysRevB.48.10345} {\bibfield  {journal} {\bibinfo
  {journal} {Phys. Rev. B}\ }\textbf {\bibinfo {volume} {48}},\ \bibinfo
  {pages} {10345} (\bibinfo {year} {1993})}\BibitemShut {NoStop}%
\bibitem [{\citenamefont {Gebhard}\ \emph {et~al.}(2022)\citenamefont
  {Gebhard}, \citenamefont {Bauerbach},\ and\ \citenamefont
  {Legeza}}]{Gebhard2022}%
  \BibitemOpen
  \bibfield  {author} {\bibinfo {author} {\bibfnamefont {F.}~\bibnamefont
  {Gebhard}}, \bibinfo {author} {\bibfnamefont {K.}~\bibnamefont {Bauerbach}},\
  and\ \bibinfo {author} {\bibfnamefont {{\"{O}}.}~\bibnamefont {Legeza}},\
  }\bibfield  {title} {\bibinfo {title} {{Accurate localization of
  Kosterlitz-Thouless-type quantum phase transitions for one-dimensional
  spinless fermions}},\ }\href {https://doi.org/10.1103/PhysRevB.106.205133}
  {\bibfield  {journal} {\bibinfo  {journal} {Physical Review B}\ }\textbf
  {\bibinfo {volume} {106}},\ \bibinfo {pages} {205133} (\bibinfo {year}
  {2022})}\BibitemShut {NoStop}%
\bibitem [{\citenamefont {Wessel}\ \emph {et~al.}(2017)\citenamefont {Wessel},
  \citenamefont {Normand}, \citenamefont {Mila},\ and\ \citenamefont
  {Honecker}}]{Wessel2017}%
  \BibitemOpen
  \bibfield  {author} {\bibinfo {author} {\bibfnamefont {S.}~\bibnamefont
  {Wessel}}, \bibinfo {author} {\bibfnamefont {B.}~\bibnamefont {Normand}},
  \bibinfo {author} {\bibfnamefont {F.}~\bibnamefont {Mila}},\ and\ \bibinfo
  {author} {\bibfnamefont {A.}~\bibnamefont {Honecker}},\ }\bibfield  {title}
  {\bibinfo {title} {{Efficient Quantum Monte Carlo simulations of highly
  frustrated magnets: the frustrated spin-1/2 ladder}},\ }\href
  {https://doi.org/10.21468/SciPostPhys.3.1.005} {\bibfield  {journal}
  {\bibinfo  {journal} {SciPost Physics}\ }\textbf {\bibinfo {volume} {3}},\
  \bibinfo {pages} {005} (\bibinfo {year} {2017})}\BibitemShut {NoStop}%
\bibitem [{\citenamefont {Bauer}\ \emph {et~al.}(2011)\citenamefont {Bauer},
  \citenamefont {Carr}, \citenamefont {Evertz}, \citenamefont {Feiguin},
  \citenamefont {Freire}, \citenamefont {Fuchs}, \citenamefont {Gamper},
  \citenamefont {Gukelberger}, \citenamefont {Gull}, \citenamefont {Guertler},
  \citenamefont {Hehn}, \citenamefont {Igarashi}, \citenamefont {Isakov},
  \citenamefont {Koop}, \citenamefont {Ma}, \citenamefont {Mates},
  \citenamefont {Matsuo}, \citenamefont {Parcollet}, \citenamefont
  {Paw{\l}owski}, \citenamefont {Picon}, \citenamefont {Pollet}, \citenamefont
  {Santos}, \citenamefont {Scarola}, \citenamefont {Schollw{\"{o}}ck},
  \citenamefont {Silva}, \citenamefont {Surer}, \citenamefont {Todo},
  \citenamefont {Trebst}, \citenamefont {Troyer}, \citenamefont {Wall},
  \citenamefont {Werner},\ and\ \citenamefont {Wessel}}]{Bauer2011}%
  \BibitemOpen
  \bibfield  {author} {\bibinfo {author} {\bibfnamefont {B.}~\bibnamefont
  {Bauer}}, \bibinfo {author} {\bibfnamefont {L.~D.}\ \bibnamefont {Carr}},
  \bibinfo {author} {\bibfnamefont {H.~G.}\ \bibnamefont {Evertz}}, \bibinfo
  {author} {\bibfnamefont {A.}~\bibnamefont {Feiguin}}, \bibinfo {author}
  {\bibfnamefont {J.}~\bibnamefont {Freire}}, \bibinfo {author} {\bibfnamefont
  {S.}~\bibnamefont {Fuchs}}, \bibinfo {author} {\bibfnamefont
  {L.}~\bibnamefont {Gamper}}, \bibinfo {author} {\bibfnamefont
  {J.}~\bibnamefont {Gukelberger}}, \bibinfo {author} {\bibfnamefont
  {E.}~\bibnamefont {Gull}}, \bibinfo {author} {\bibfnamefont {S.}~\bibnamefont
  {Guertler}}, \bibinfo {author} {\bibfnamefont {A.}~\bibnamefont {Hehn}},
  \bibinfo {author} {\bibfnamefont {R.}~\bibnamefont {Igarashi}}, \bibinfo
  {author} {\bibfnamefont {S.~V.}\ \bibnamefont {Isakov}}, \bibinfo {author}
  {\bibfnamefont {D.}~\bibnamefont {Koop}}, \bibinfo {author} {\bibfnamefont
  {P.~N.}\ \bibnamefont {Ma}}, \bibinfo {author} {\bibfnamefont
  {P.}~\bibnamefont {Mates}}, \bibinfo {author} {\bibfnamefont
  {H.}~\bibnamefont {Matsuo}}, \bibinfo {author} {\bibfnamefont
  {O.}~\bibnamefont {Parcollet}}, \bibinfo {author} {\bibfnamefont
  {G.}~\bibnamefont {Paw{\l}owski}}, \bibinfo {author} {\bibfnamefont {J.~D.}\
  \bibnamefont {Picon}}, \bibinfo {author} {\bibfnamefont {L.}~\bibnamefont
  {Pollet}}, \bibinfo {author} {\bibfnamefont {E.}~\bibnamefont {Santos}},
  \bibinfo {author} {\bibfnamefont {V.~W.}\ \bibnamefont {Scarola}}, \bibinfo
  {author} {\bibfnamefont {U.}~\bibnamefont {Schollw{\"{o}}ck}}, \bibinfo
  {author} {\bibfnamefont {C.}~\bibnamefont {Silva}}, \bibinfo {author}
  {\bibfnamefont {B.}~\bibnamefont {Surer}}, \bibinfo {author} {\bibfnamefont
  {S.}~\bibnamefont {Todo}}, \bibinfo {author} {\bibfnamefont {S.}~\bibnamefont
  {Trebst}}, \bibinfo {author} {\bibfnamefont {M.}~\bibnamefont {Troyer}},
  \bibinfo {author} {\bibfnamefont {M.~L.}\ \bibnamefont {Wall}}, \bibinfo
  {author} {\bibfnamefont {P.}~\bibnamefont {Werner}},\ and\ \bibinfo {author}
  {\bibfnamefont {S.}~\bibnamefont {Wessel}},\ }\bibfield  {title} {\bibinfo
  {title} {{The ALPS project release 2.0: open source software for strongly
  correlated systems}},\ }\href
  {https://doi.org/10.1088/1742-5468/2011/05/P05001} {\bibfield  {journal}
  {\bibinfo  {journal} {J. Stat. Mech.: Theory Exp.}\ }\textbf {\bibinfo
  {volume} {2011}}\bibinfo  {number} { (05)},\ \bibinfo {pages}
  {P05001}}\BibitemShut {NoStop}%
\bibitem [{\citenamefont {Fishman}\ \emph {et~al.}(2022)\citenamefont
  {Fishman}, \citenamefont {White},\ and\ \citenamefont
  {Stoudenmire}}]{SciPost}%
  \BibitemOpen
\bibfield  {number} {  }\bibfield  {author} {\bibinfo {author} {\bibfnamefont
  {M.}~\bibnamefont {Fishman}}, \bibinfo {author} {\bibfnamefont {S.~R.}\
  \bibnamefont {White}},\ and\ \bibinfo {author} {\bibfnamefont {E.~M.}\
  \bibnamefont {Stoudenmire}},\ }\bibfield  {title} {\bibinfo {title} {{The
  ITensor Software Library for Tensor Network Calculations}},\ }\href
  {https://doi.org/10.21468/SciPostPhysCodeb.4} {\bibfield  {journal} {\bibinfo
   {journal} {SciPost Phys. Codebases}\ ,\ \bibinfo {pages} {4}} (\bibinfo
  {year} {2022})}\BibitemShut {NoStop}%
\bibitem [{\citenamefont {Oshikawa}\ \emph {et~al.}(1997)\citenamefont
  {Oshikawa}, \citenamefont {Yamanaka},\ and\ \citenamefont
  {Affleck}}]{oshikawa}%
  \BibitemOpen
  \bibfield  {author} {\bibinfo {author} {\bibfnamefont {M.}~\bibnamefont
  {Oshikawa}}, \bibinfo {author} {\bibfnamefont {M.}~\bibnamefont {Yamanaka}},\
  and\ \bibinfo {author} {\bibfnamefont {I.}~\bibnamefont {Affleck}},\
  }\bibfield  {title} {\bibinfo {title} {Magnetization plateaus in spin chains:
  ``haldane gap'' for half-integer spins},\ }\href
  {https://doi.org/10.1103/PhysRevLett.78.1984} {\bibfield  {journal} {\bibinfo
   {journal} {Phys. Rev. Lett.}\ }\textbf {\bibinfo {volume} {78}},\ \bibinfo
  {pages} {1984} (\bibinfo {year} {1997})}\BibitemShut {NoStop}%
\bibitem [{\citenamefont {Cazalilla}\ \emph {et~al.}(2011)\citenamefont
  {Cazalilla}, \citenamefont {Citro}, \citenamefont {Giamarchi}, \citenamefont
  {Orignac},\ and\ \citenamefont {Rigol}}]{Cazalilla_2011}%
  \BibitemOpen
  \bibfield  {author} {\bibinfo {author} {\bibfnamefont {M.~A.}\ \bibnamefont
  {Cazalilla}}, \bibinfo {author} {\bibfnamefont {R.}~\bibnamefont {Citro}},
  \bibinfo {author} {\bibfnamefont {T.}~\bibnamefont {Giamarchi}}, \bibinfo
  {author} {\bibfnamefont {E.}~\bibnamefont {Orignac}},\ and\ \bibinfo {author}
  {\bibfnamefont {M.}~\bibnamefont {Rigol}},\ }\bibfield  {title} {\bibinfo
  {title} {One dimensional bosons: From condensed matter systems to ultracold
  gases},\ }\href {https://doi.org/10.1103/RevModPhys.83.1405} {\bibfield
  {journal} {\bibinfo  {journal} {Rev. Mod. Phys.}\ }\textbf {\bibinfo {volume}
  {83}},\ \bibinfo {pages} {1405} (\bibinfo {year} {2011})}\BibitemShut
  {NoStop}%
\bibitem [{\citenamefont {Zapf}\ \emph {et~al.}(2014)\citenamefont {Zapf},
  \citenamefont {Jaime},\ and\ \citenamefont {Batista}}]{Zapf2014}%
  \BibitemOpen
  \bibfield  {author} {\bibinfo {author} {\bibfnamefont {V.}~\bibnamefont
  {Zapf}}, \bibinfo {author} {\bibfnamefont {M.}~\bibnamefont {Jaime}},\ and\
  \bibinfo {author} {\bibfnamefont {C.~D.}\ \bibnamefont {Batista}},\
  }\bibfield  {title} {\bibinfo {title} {Bose-einstein condensation in quantum
  magnets},\ }\href {https://doi.org/10.1103/RevModPhys.86.563} {\bibfield
  {journal} {\bibinfo  {journal} {Reviews of Modern Physics}\ }\textbf
  {\bibinfo {volume} {86}},\ \bibinfo {pages} {563} (\bibinfo {year}
  {2014})}\BibitemShut {NoStop}%
\bibitem [{\citenamefont {Kühner}\ \emph {et~al.}(2000)\citenamefont
  {Kühner}, \citenamefont {White},\ and\ \citenamefont {Monien}}]{Kuhner}%
  \BibitemOpen
  \bibfield  {author} {\bibinfo {author} {\bibfnamefont {T.~D.}\ \bibnamefont
  {Kühner}}, \bibinfo {author} {\bibfnamefont {S.~R.}\ \bibnamefont {White}},\
  and\ \bibinfo {author} {\bibfnamefont {H.}~\bibnamefont {Monien}},\
  }\bibfield  {title} {\bibinfo {title} {One-dimensional bose-hubbard model
  with nearest-neighbor interaction},\ }\href
  {https://doi.org/10.1103/PhysRevB.61.12474} {\bibfield  {journal} {\bibinfo
  {journal} {Physical Review B}\ }\textbf {\bibinfo {volume} {61}},\ \bibinfo
  {pages} {12474} (\bibinfo {year} {2000})}\BibitemShut {NoStop}%
\bibitem [{\citenamefont {Jiang}\ \emph {et~al.}(2023)\citenamefont {Jiang},
  \citenamefont {Romhányi}, \citenamefont {White}, \citenamefont
  {Zhitomirsky},\ and\ \citenamefont {Chernyshev}}]{Jiang2023}%
  \BibitemOpen
  \bibfield  {author} {\bibinfo {author} {\bibfnamefont {S.}~\bibnamefont
  {Jiang}}, \bibinfo {author} {\bibfnamefont {J.}~\bibnamefont {Romhányi}},
  \bibinfo {author} {\bibfnamefont {S.~R.}\ \bibnamefont {White}}, \bibinfo
  {author} {\bibfnamefont {M.}~\bibnamefont {Zhitomirsky}},\ and\ \bibinfo
  {author} {\bibfnamefont {A.}~\bibnamefont {Chernyshev}},\ }\bibfield  {title}
  {\bibinfo {title} {Where is the quantum spin nematic?},\ }\href
  {https://doi.org/10.1103/PhysRevLett.130.116701} {\bibfield  {journal}
  {\bibinfo  {journal} {Physical Review Letters}\ }\textbf {\bibinfo {volume}
  {130}},\ \bibinfo {pages} {116701} (\bibinfo {year} {2023})}\BibitemShut
  {NoStop}%
\end{thebibliography}
\end{document}